\documentclass[prd,twocolumn,amsmath,amssymb,floatfix,superscriptaddress,nofootinbib]{revtex4-1}

\usepackage{graphicx}
\usepackage{scrextend}
\usepackage{amssymb}
\usepackage{amsmath}
\usepackage{bm}
\usepackage{color}
\usepackage{multirow}
\usepackage{cprotect}

\DeclareMathAlphabet\mathbfcal{OMS}{cmsy}{b}{n}
\definecolor{darkgreen}{RGB}{50,150,0}
\definecolor{purple}{cmyk}{0.5,1.0,0,0}

\def\edth{\;\raise1.0pt\hbox{$'$}\hskip-6pt\partial}
\def\baredth{\;\overline{\raise1.0pt\hbox{$'$}\hskip-6pt
\partial}}

\def\be{\begin{equation}}
\def\ee{\end{equation}}
\def\ben{\begin{equation} \nonumber}
\def\een{\end{equation}}
\def\ban{\begin{eqnarray*}}
\def\ean{\end{eqnarray*}}
\def\ba{\begin{eqnarray}}
\def\ea{\end{eqnarray}}
\def\({\left(}
\def\){\right)}

\newcommand{\Msun}{M_\odot}

\definecolor{ultramarine}{rgb}{0.07, 0.04, 0.56}
\definecolor{cadmiumgreen}{rgb}{0.0, 0.42, 0.24}
\definecolor{indigo(dye)}{rgb}{0.0, 0.25, 0.42}
\usepackage[linktocpage=true]{hyperref}
\hypersetup{
colorlinks=true,
citecolor=ultramarine,
linkcolor=cadmiumgreen,
urlcolor=indigo(dye),
pdfauthor={},
pdftitle={},
pdfsubject={}
}

\begin{document}

\title{Correlating galaxy shapes and initial conditions: an observational study}

\author{Pavel Motloch}
\affiliation{Canadian Institute for Theoretical Astrophysics, University of Toronto, M5S 3H8, ON, Canada}

\author{Ue-Li~Pen}
\affiliation{Canadian Institute for Theoretical Astrophysics, University of Toronto, M5S 3H8, ON, Canada}
\affiliation{Department of Physics, University of Toronto, 60 St. George Street, Toronto, ON M5S 1A7, Canada}
\affiliation{Institute of Astronomy and Astrophysics, Academia Sinica,
Astronomy-Mathematics Building, No. 1, Section 4, Roosevelt Road, Taipei 10617, Taiwan}
\affiliation{Perimeter Institute for Theoretical Physics, Waterloo, N2L 2Y5, ON, Canada}
\affiliation{Canadian Institute for Advanced Research, CIFAR Program in Gravitation and Cosmology, Toronto, M5G 1Z8, ON, Canada}
\affiliation{Dunlap Institute for Astronomy and Astrophysics, University of Toronto, M5S 3H4, ON, Canada}

\author{Hao-Ran~Yu}
\affiliation{Department of Astronomy, Xiamen University, Xiamen, Fujian 361005, China}
\begin{abstract}
\noindent
Using data from the Sloan Digital Sky Survey we study correlations between 
directions of galaxy angular momenta determined from images of spiral galaxies and various
observables derived from the reconstructed initial conditions. We find an apparent
systematic effect consistent with galaxy-orientation-dependent selection function. After
restricting our attention to the brightest half of the galaxies where this systematic
effect is presumed to be absent, we find hints of excess/deficit correlation for two
observables.  Interestingly, tidal torque theory predicts excess/deficit correlation in exactly
these two observables. After correcting for the redshift space distortions, the
significance of these correlations drops below 3$\sigma$ threshold. We do not find
any other systematic issues, but a thorough systematic analysis goes beyond the scope of
this work.
\end{abstract}

\maketitle

\section{Introduction}
\label{sec:intro}

Galaxy surveys have been instrumental in enhancing our understanding of the Universe
\cite{Geller:1989da, Colless:2003wz, Driver:2010zb,2000AJ....120.1579Y}. Beyond the
traditional analyses, it is useful to think about how to extract additional information
from the large number of galaxies that will be discovered in the following years
\cite{Levi:2013gra,LSSTScience:2009jmu}.

One of the suggestions has been using measurements of galaxy angular momenta, that can
extend our understanding of physics at megaparsec scales
\cite{Lee:1999ii,Lee:2000br,Yu:2019bsd}. Such measurements can aid with reconstruction of the
initial conditions (ICs) in the Universe \cite{Lee:1999ii,Lee:2000br} and help to
constrain neutrino masses \cite{Yu:2018llx}, primordial gravitational waves 
\cite{Biagetti:2020lpx}, primordial non-Gaussianity \cite{Schmidt:2015xka}
and chirality violations in the early Universe \cite{Yu:2019bsd}. Better knowledge of
initial matter distribution in the local volume would also aid studies of galaxy
formation.

In our current understanding, the dark matter haloes acquire angular momenta from the
inhomogeneous tidal field that torques the non-spherical protohalo early on
\cite{Peebles:1969jm,1970Ap......6..320D,1984ApJ...286...38W,
Porciani:2001db,Porciani:2001er}.
At late times, interactions
with the nearby large scale structure notably complicate the picture
\cite{Porciani:2001db,Porciani:2001er, Krolewski:2019bfv,
2017ApJ...841...16D,Zhang:2014rju,Veena:2018ooo,Veena:2019ozd,Kraljic:2021oeg,Kraljic:2019lca,
Wang:2017tsr,Neyrinck:2019uvc,Schaefer:2008xd,Codis:2012ep,Codis:2015tla,Welker:2019puz,
Jones:2010cs,AragonCalvo:2006ay,Hahn:2007ui,Bett:2011rs,Bett:2015aoa,Tempel:2013gqa,Wang:2018rlf}.

Despite these late time effects, it turns out that the {\emph {direction} } of the halo
angular momentum retains a significant amount of the initial information
\cite{Yu:2019bsd}.
Naturally, we can not directly measure angular momenta of dark matter haloes.
Fortunately, spins\footnote{
As a shorthand, here and in what follows we use ``spin'' to specifically only refer to the
direction of the angular momentum of a halo/galaxy, ignoring its amplitude.
} of galaxies tend to be correlated with spins of their dark matter haloes
\cite{2015ApJ...812...29T, Jiang:2018ioo},
and so there is a potential to use measurements of galaxy spins to probe the initial conditions
in the local Universe.

It seems that the {\emph{amplitude}} of the dark matter halo's angular momentum
also retains information about the initial conditions \cite{Wu:2020btz} (but see
\cite{Porciani:2001db,Jiang:2018ioo}). However, in this work we only consider the angular
momenta directions, especially given the experimental cost involved in determining the
amplitudes of galaxy angular momenta.

The prospect that measurements of galaxy spins might help with the reconstruction of the
initial conditions in the Universe requires an observational verification. 
We made a first step in this
direction in \cite{Motloch:2020qvx}, where we found a correlation between the galaxy spins of
a sample of about 15000 galaxies and initial conditions reconstructed from
positions of a larger sample of galaxies \cite{Wang:2016qbz}. Unfortunately, due to the
limited number of galaxies we could only confirm this correlation with an $\sim 3 \sigma$
confidence.
To achieve this result, we combined a set of galaxies for which integral field
spectroscopy data are available with a set of spiral galaxies with known senses of rotation
of their spiral arms.

At least in principle we have access to a significantly larger dataset of galaxy spins.
Assuming that each spiral galaxy forms a thin disk with its spin pointing perpendicularly to
the plane of the disk, we can utilize shape measurements of spiral galaxies to determine
their spins up to a so-called four-fold
degeneracy (see below) \cite{Lee:1999ii}. Shape
measurements of spiral galaxies, as a proxy for their spins, are thus potentially usable
as an additional handle on the initial conditions of the Universe. 
In this paper we investigate to what extent such a proposition is currently viable by studying
correlations between galaxy spin components determined solely from galaxy shapes and
observables derived from the initial density field as measured using a traditional reconstruction
technique. In contrast with \cite{Motloch:2020qvx}, we trade a somewhat lower
signal per galaxy (due to the four-fold degeneracy) for a larger number of galaxies.
Unlike
the previous studies such as \cite{Lee:2007jx}, we study correlations of
galaxy shapes with observables derived from the \emph{initial}, not late time,
density field.

This paper is organized as follows:
In \S~\ref{sec:theory} we describe how to determine galaxy spins from the galaxy shape
measurements and discuss the four-fold degeneracy. Then we explain how we quantify correlations
between galaxy spins and the initial conditions and list the observables built from the
initial conditions we consider in this work. In \S~\ref{sec:data} we introduce the data
used and in \S~\ref{sec:results} we present our results. We conclude with discussion in
\S~\ref{sec:discuss}.

We use indices $i, j, k, l$ to denote components of vectors and tensors in three
dimensions.

\section{Theory}
\label{sec:theory}

In this section we introduce the thin disk approximation that we use to connect
measurements of galaxy shapes and their angular momenta. Then we describe how we
quantify correlations between galaxy spins and various observables built from the initial
conditions, before listing all such observables considered in this work.

\subsection{Spins from shapes}

Measuring the full 3D vector of galaxy angular momentum requires integral field
spectroscopy, which is experimentally costly and can at the present only be obtained for
several thousands of galaxies \cite{Aguado:2018ynx,2018MNRAS.481.2299S}. Using an approximation in which
spiral galaxies are considered to be thin disks allows us to use images of spiral galaxies
to infer information about their
angular momenta \cite{Lee:2007jx, Kraljic:2021oeg}. Currently, tens of thousands of
such images are available. 

In this thin-disk approximation, matter is assumed to be circling around the galaxy center in the
plane of the disk, implying that the galaxy angular momentum vector is perpendicular to
the plane of the galaxy. This plane can be, up to a two-fold degeneracy, determined from
the position angle $\alpha$ (measured relative to the north celestial pole, turning
positive into the direction of the right ascension) and axis ratio $\mathcal{R}$ of the galaxy image. Additional
two-fold degeneracy then arises because from $\alpha$ and $\mathcal{R}$ alone we are
unable to determine whether the galaxy spin points towards or away from the observer.

The unit spin vector of such a galaxy in the local spherical coordinate frame $(L_r,
L_\theta, L_\phi)$ can be determined through \cite{Pen:2000ug}
\ba
\label{spherical_L}
  |L_r| &=& \mathcal{R}\nonumber\\
  L_\theta/L_\phi &=& \tan \alpha ,
\ea
where the four-fold degeneracy prevents us from determining the signs of $L_r$ and
$L_\theta$ (and thus $L_\phi$) from the galaxy image alone.

In the equatorial Cartesian coordinates, the unit spin vector can be then determined from
the right ascension and declination of the galaxy through \cite{Lee:2007jx}
\ba
\label{cartesian_L}
  \vec L = \vec L_R + \vec L_T ,
\ea
where we defined
\ba
\label{LR_LT}
  \vec L_R &\propto& L_r \(
   \sin \theta \cos \phi,
  \sin \theta \sin \phi,
  \cos \theta
  \)
  \nonumber\\
  \vec {L}_{T} &\propto& 
  L_\theta \(
  \cos \theta \cos \phi,
  \cos \theta \sin \phi,
  - \sin \theta 
  \) 
  \nonumber\\
  && + L_\phi \(
  - \sin \phi,
  \cos \phi, 0
  \) 
\ea
and $\theta = \pi/2 - \mathrm{DEC}$ and $\phi = \mathrm{RA}$.
Because of the four-fold degeneracy, signs of $\vec L_{R,T}$ are ambiguous.
Notice that in general neither of the vectors $\vec L_{R,T}$ has unit norm.

By default we consider spiral galaxies to be infinitesimally thin disks and use
\eqref{spherical_L}. To test for systematic effect of this assumption, we also consider
non-zero intrinsic flatness $p$ of the spiral galaxies, where we replace
$\mathcal{R}$ in \eqref{spherical_L} with
\be
\label{intrinsic_flatness_replacement}
  \mathcal{R} \rightarrow \sqrt{\frac{\mathrm{max}\(\mathcal{R}^2 - p^2,0\)}{1 - p^2}} .
\ee
To derive this equation, one assumes that spiral galaxy is an ellipsoid with lengths of
the principal axes in the ratio 1:1:$p$. Then the observed projected axis ratio
$\mathcal{R}$ is related to the galaxy inclination angle $i$ according to
\be
  \mathcal{R} = \sqrt{\cos^2 i + p^2 \sin^2 i} , 
\ee
from which \eqref{intrinsic_flatness_replacement} follows given that \eqref{spherical_L} is
just a special case of the more general
\be
  |L_r| = \cos i .
\ee
The maximum in \eqref{intrinsic_flatness_replacement} just enforces that any galaxy with
$\mathcal{R}$ below $p$ is considered to have inclination angle $i = \pi/2$.
Experimentally, $p$ of spiral galaxies has been measured to be $\sim 0.1 - 0.2$
\cite{1984AJ.....89..758H}. 

\subsection{Quantifying correlations}

In this section we give an overview of how we quantify correlations between
measurements of galaxy spins as represented by $\vec L_{R,T}$ and the initial conditions. In
all cases, the initial conditions are represented by either a vector $V_i$ or a
rank-two tensor $M_{ij}$ constructed from the second derivatives of the initial density
$\rho_\mathrm{ini}$, gravitational potential $\phi_\mathrm{ini}$, or both. See the next
section for the list of $V_i$ and $M_{ij}$ used in this work

Considering first the rank-two variables of the form $M_{ij}$, we can form three scalar
combinations
\ba
  L_R M L_R &\equiv& \sum_{ij} L_{Ri} M_{ij} L_{Rj}\\
  L_T M L_T &\equiv& \sum_{ij} L_{Ti} M_{ij} L_{Tj}\\
  L_T M L_R &\equiv& \sum_{ij} L_{Ti} M_{ij} L_{Rj} ,
\ea
where for future convenience we suppressed the index structure on the left hand side.
Given the four-fold degeneracy in determining $\vec L_{T}$ and $\vec L_{R}$, the last
combination can not be evaluated using the shape data alone. On the other hand,
$L_RML_R$ and $L_T M L_T$ are invariant with respect to this degeneracy and can be used to
study the correlations between galaxy spins and initial conditions.

With vector variables $V_i$ the situation is slightly more complicated, as the naive
combinations $\vec L_{R,T} \cdot \vec V$ are again indeterminate due to the
uncertainty in the sign of $\vec L_{R,T}$. We can form an invariant quantity by taking the
absolute value of the scalar product,
\ba
  \left|L_R V\right| &\equiv& \left|\sum_{i} L_{Ri} V_{i}\right|\\
  \left|L_T V\right| &\equiv& \left|\sum_{i} L_{Ti} V_{i}\right| ,
\ea
which can be used to study the correlation of $\vec L_{R,T}$ with the initial conditions.
Notice that more information is available when we are able to break the
degeneracy \cite{Motloch:2020qvx}.

For each variable $\mathcal{V}$ of the form $L_\alpha M L_\alpha$ or $\left| L_\alpha V
\right|$  with $\alpha \in \{R,T\}$ we then
evaluate the average $\langle \mathcal{V} \rangle_\mathrm{data}$ over the galaxies in our sample. To
determine the probability of observing such $\langle\mathcal{V}\rangle_\mathrm{data}$ in case of
uncorrelated spins $\vec L_{R,T}$ and initial conditions (as represented by $M_{ij}$ /
$V_i$), we calculate $\mathcal{V}$ for many mock galaxy catalogs in which the
galaxy positions are kept fixed but we randomly shuffle the parameters $\alpha, \mathcal{R}$ that
determine the galaxy spins $\vec L_{R,T}$. As a result, we get the mean and standard
deviation $\langle{\mathcal{V}}\rangle_\mathrm{rnd}, \sigma_{\langle\mathcal{V}\rangle}$
of this random mock distribution\footnote{But notice this procedure can give a slightly underestimated standard
deviation, see for example \cite{Friedrich:2015nga}. }, which allows us to calculate the statistical
significance of the detected excess correlation
\be
  S = \frac{\left|\langle \mathcal{V}\rangle_\mathrm{data} -
  \langle{\mathcal{V}}\rangle_\mathrm{rnd}\right|}{\sigma_{\langle\mathcal{V}\rangle}} .
\ee
Notice it is incorrect to shuffle the vectors $\vec L_{R,T}$ themselves, because then
these vectors would no longer be parallel to / perpendicular to the line of sight.

\subsection{Functions of initial conditions}

In this section we introduce variables $M_{ij},V_i$ we use to characterize the initial conditions
and which we correlate with the galaxy spins described by $\vec
L_{R,T}$. 

It has been shown in both simulations \cite{Yu:2019bsd} and observations
\cite{Motloch:2020qvx} that galaxy spins correlate with the vector field $\vec
L^\mathrm{IC}$ defined in terms of the smoothed initial density $\rho_\mathrm{ini}^r$ and gravitational
potential $\phi_\mathrm{ini}^r$ as
\be
\label{limit_formula}
L^\mathrm{IC}_i
=
\sum_{jkl}
\epsilon_{ijk}\,
\(\partial_{jl} \phi_\mathrm{ini}^r\)\,
\partial_{lk} \rho_\mathrm{ini}^r .
\ee
This vector field is a straightforward extension of the standard tidal torque theory
formula \cite{1984ApJ...286...38W}
\be
L^\mathrm{TTT}_i \propto
\sum_{jkl}
\epsilon_{ijk}\,
\(\partial_{jl} \phi\)\,
I_{lk} ,
\ee
where $I_{lk}$ is the moment of inertia tensor of the protohalo.
In \eqref{limit_formula}, the fields have been
smoothed with a Gaussian kernel with smoothing scale $r \sim O(h^{-1}\mathrm{Mpc})$;
this is represented by the
superscript $r$. The unit norm vector field $\hat
L^\mathrm{IC} = \vec L^\mathrm{IC}/|L^\mathrm{IC}|$ is then the first observable built
from the initial
conditions we correlate with the galaxy spins.

It has long been understood that tidal field in the vicinity of a protohalo plays a
crucial role in determining the halo's spin \cite{Peebles:1969jm, 1970Ap......6..320D,
1984ApJ...286...38W, Porciani:2001db,Porciani:2001er}. Tidal torque theory suggests that
galaxy spins should correlate with the second power of the unit traceless local shear
tensor $\hat T$ \cite{Lee:2000br}. This tensor is built from the shear tensor,
defined as 
\be
\label{shear_definition}
  T_{ij} = \partial_{ij} \phi_\mathrm{ini}^r ,
\ee
by first subtracting the trace
\be
  \tilde T_{ij} = T_{ij} - \frac{\delta_{ij}}{3} \sum_k T_{kk}
\ee
and then normalizing
\be
  \hat T_{ij} = \frac{\tilde T_{ij}}{\sqrt{\sum_{ij} \tilde T_{ij}^2}} .
\ee
This tidal torque theory prediction motivates us to take $(\hat T)^2$ as our second variable.

The tidal-torque theory also predicts that the correlation $\langle L_\alpha \hat T
L_\alpha\rangle$ should vanish \cite{Lee:2000br}: Assuming Gaussian initial conditions, initial
perturbation $\phi_\mathrm{ini}$ and its reverse $- \phi_\mathrm{ini}$ are equally likely.
However, the observable $\langle L_\alpha \hat T L_\alpha\rangle$ is not invariant under the
operation $\phi_\mathrm{ini} \rightarrow - \phi_\mathrm{ini}$ as $\hat T_{ij}$ changes
sign but the quadratic form $L_{\alpha,i}L_{\alpha,j}$ does not. Physically, tidal-torque theory
predicts equal alignments of galaxy spins with major and minor principal axes of $\hat T$,
which leads to cancellations in $\langle L_\alpha \hat T L_\alpha\rangle$. However,
various studies suggest that tidal-torque theory provides an incomplete picture of how the 
galaxy angular momenta arise \cite{Porciani:2001db, Hui:2002xt}. Because of this, we also
consider correlations with the first power of $\hat T$.

One can expect similar sensitivity to the second derivative of the initial
density field, which can serve as a proxy for the protohalo's moment of inertia
\cite{Motloch:2020qvx, Codis:2015tla}. We thus also investigate correlations with first
and second power of $\hat I_{ij}$, defined similarly to $\hat T_{ij}$ but with
$\phi_\mathrm{ini}^r$ in \eqref{shear_definition} replaced by $\rho_\mathrm{ini}^r$.

Finally, tidal-torque theory provides theoretical arguments that the galaxy spins should
preferentially orient with the intermediate principal axis $\hat L^\phi_0$ of the shear tensor
$\hat T$ \cite{Lee:1999ii}. In the limit where the tensor $\hat I$ can be used as a proxy
for the protohalo moment of inertia, the same arguments can be used to justify expectation of
correlation of galaxy spins with the intermediate principal axis $\hat L^\rho_0$ of $\hat
I$. The two unit vectors $\hat L^{\phi,\rho}_0$ are then the last two observables built from the
initial conditions that we consider in this work.

Overall, we study correlations with seven objects constructed from the initial conditions:
four matrices $\hat T, \(\hat T\)^2, \hat I, \(\hat I\)^2$ and three vectors $\hat L^\phi_0,
\hat L^\rho_0, \hat L^\mathrm{IC}$. We correlate them with either $\vec L_R$ or $\vec L_T$, which
leads to 14 combinations to study for each of the smoothing scales we consider.
Because we talk about the initial
conditions, all these variables are evaluated at the {\emph{Lagrangian}} positions of the
galaxies.

\section{Data}
\label{sec:data}

In this section we present the data used in this work. We start by describing the initial
conditions as reconstructed by the ELUCID Collaboration and then introduce the data used
to determine angular momenta and positions of galaxies.

\subsection{Initial conditions}

The initial density field $\rho_\mathrm{ini}$ used in this work was obtained by the ELUCID
Collaboration. Extended details about their methodology can be found in
\cite{2014ApJ...794...94W,Wang:2016qbz}.

They first created a catalog of galaxy groups from the Sloan Digital Sky Survey (SDSS) data
\cite{Yang:2007yr} and then estimated mass of each group via a luminosity-based
abundance matching. After correcting for peculiar velocities, they only retained groups in the
Northern Galactic Cap, redshift range $0.01 \le z \le 0.12$ and with masses above
$10^{12}\Msun$. The space was then tessellated
according to which galaxy group was the closest. Within the resulting sub-volumes,
particles were
placed randomly, in accordance with the expected density profile for halo of given mass.
This particle distribution represents today's density field.

In the second step of the reconstruction, ELUCID Collaboration determined the
best fit initial conditions by repeatedly running a Particle-Mesh (PM)
dynamics code in a Hamiltonian Monte Carlo fashion. For each random set of initial
conditions, the PM code was used to calculate the corresponding value of today's density
field. This density field was compared with that determined from the SDSS data and their
relative closeness was quantified using a predefined measure. As the
PM code is inaccurate on small scales, both density fields were smoothed
with a Gaussian kernel with smoothing scale
$4\, \mathrm{Mpc/h}$ before comparison. Iteratively probing the space of initial
conditions then allowed ELUCID to find the initial conditions that best describe the local
galaxy data.

From the best fit $\rho_\mathrm{ini}$, we use the Poisson equation to calculate the
initial gravitational potential $\phi_\mathrm{ini}$.

In \cite{Motloch:2020qvx} we found that with ELUCID ICs, galaxy spins are best predicted
when smoothing the initial conditions with $r \sim 3\, \mathrm{Mpc/h}$. In this work we
consider this smoothing scale, together with $r \sim 2\, \mathrm{Mpc/h}$ and
$r \sim 4\, \mathrm{Mpc/h}$ to allow for the possibility of a potentially different optimal
smoothing scale for the observables studied here.

\subsection{Galaxy shapes and positions}

Galaxy data used in this work were obtained by SDSS \cite{2000AJ....120.1579Y,
Aguado:2018ynx} and we downloaded them from
CasJobs\footnote{https://skyserver.sdss.org/CasJobs/}. First we use
the Galaxy Zoo classifications \cite{Lintott:2008ne} to select only spiral galaxies.
For each such galaxy we obtain its right ascension, declination, redshift, position
angle and axis ratio.
As our default combination, for the position angle and axis ratio we use results of the
fit of the exponential profile to the galaxy images obtained using the green SDSS
filter. For systematic checks we also investigate other choices. As in our previous work,
we only consider galaxies for which the closest ELUCID halo was at least $10^{12}
M_\odot$ to avoid regions with poorly determined ICs.

For each galaxy, we calculate the vectors $\vec L_{R,T}$ according to \eqref{LR_LT}.
The redshift and sky position then allows us to find each galaxy's three-dimensional
Euclidean position. Using the simulation run from the optimal ICs, we then find the
galaxy's inverse displacement, which we use to calculate the Lagrangian position of the
galaxy.

Our full fiducial catalog contains 50361 galaxies in the volume in which ELUCID provides
the initial conditions.

\section{Results}
\label{sec:results}

In this section we first present measured excess correlations for all investigated observables when
using the full sample of galaxies. Then we discuss an apparent systematic bias
present in our results, possibly related to a galaxy-orientation-dependent selection
function. After restricting our attention to the brighter half of galaxies, we find that
only two observables show $\sim$ 4$\sigma$ hints of an excess correlation.
In the final part of this section we perform several other systematic checks
on these two observables.

\subsection{Using full galaxy catalog}

\begin{figure*}
\center
\includegraphics[width = 0.99 \textwidth]{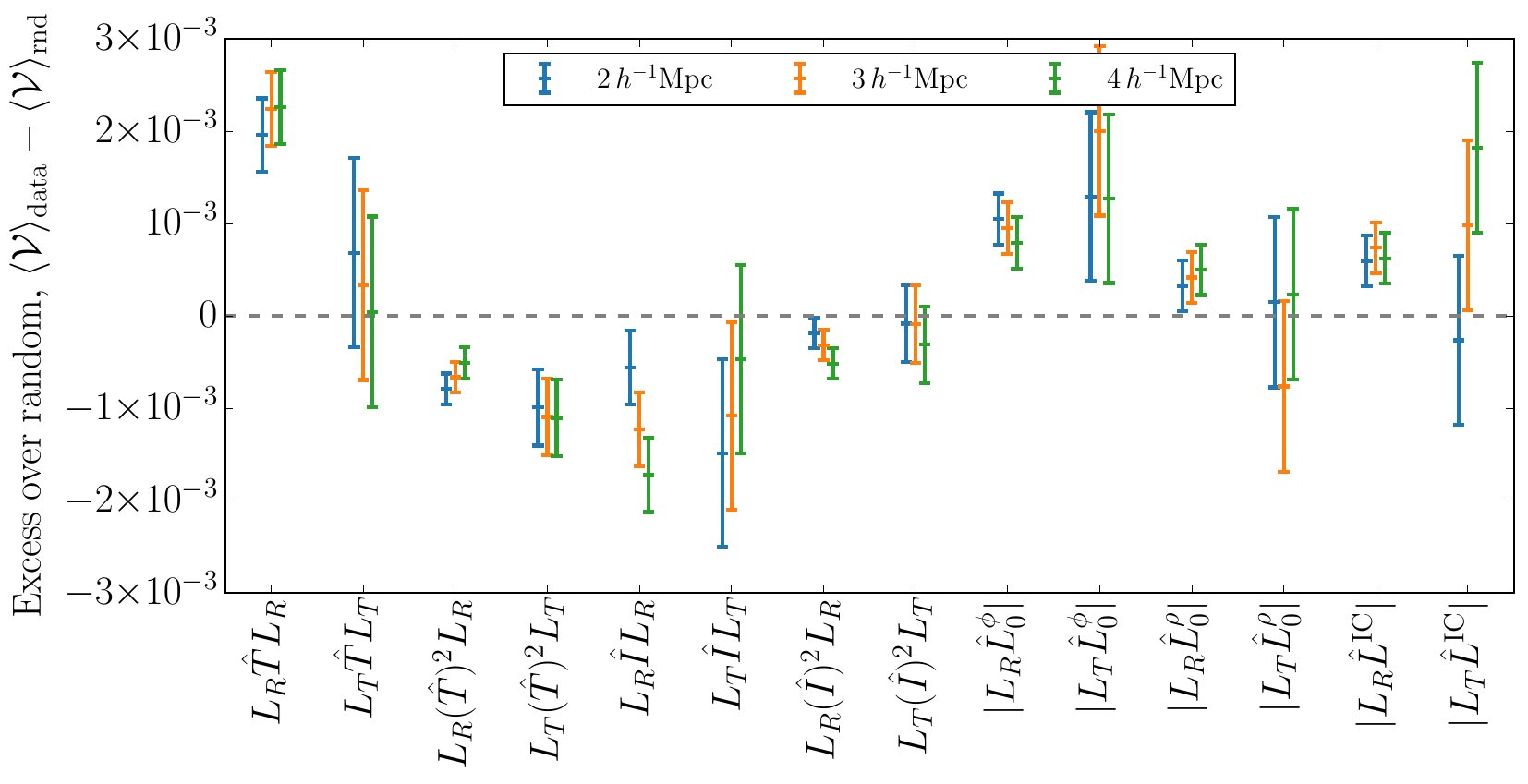}
\caption{Excess correlations of galaxy spins with various vectors and tensors derived from
the initial conditions when using the full galaxy sample. Different colors represent
different smoothing scales $r$ used to calculate the vectors and tensors.}
\label{fig:MV}
\end{figure*}

In Fig.~\ref{fig:MV} we show excess correlation between the galaxy spins $\vec L_{R,T}$
inferred from the galaxy shapes and various observables built from the initial conditions
when we use the full galaxy sample. We
summarize the detection significances $S$ in Table~\ref{tab:sigmas}. To
estimate the error bars, we shuffled $\mathcal{R}$ and $\alpha$ total of 40000 times,
which is enough to converge the results to below 0.05 standard deviation. 

\begin{table}
\caption{Significances $S$ of excess correlations in terms of number of standard
deviations when using the full galaxy sample.
}
\label{tab:sigmas}
\begin{center}
\begin{tabular}{cccc}
\hline\hline
 & \multicolumn{3}{c}{Smoothing scale $r$} \\
 & $2\, h^{-1}\mathrm{Mpc}$ & $3\, h^{-1}\mathrm{Mpc}$ & $4\, h^{-1}\mathrm{Mpc}$\\
\hline
$L_R\hat TL_R$	    &	4.9	&	5.6	&	5.7	\\
$L_T\hat TL_T$	    &	0.7	&	0.3	&	0.0	\\
$L_R(\hat T)^2L_R$	&	4.7	&	4.0	&	3.0	\\
$L_T(\hat T)^2L_T$	&	2.4	&	2.6	&	2.7	\\
$L_R\hat IL_R$	    &	1.4	&	3.1	&	4.3	\\
$L_T\hat IL_T$	    &	1.5	&	1.1	&	0.5	\\
$L_R(\hat I)^2L_R$	&	1.1	&	1.9	&	3.2	\\
$L_T(\hat I)^2L_T$	&	0.2	&	0.2	&	0.8	\\
$|L_RL_0^{\phi}|$	    &	3.8	&	3.4	&	2.9	\\
$|L_TL_0^{\phi}|$	    &	1.4	&	2.2	&	1.4	\\
$|L_RL_0^{\rho}|$	    &	1.2	&	1.5	&	1.8	\\
$|L_TL_0^{\rho}|$	    &	0.2	&	0.8	&	0.3	\\
$|L_RL^{IC}|$	      &	2.2	&	2.7	&	2.3	\\
$|L_TL^{IC}|$	      &	0.3	&	1.1	&	2.0	\\
\hline\hline
\end{tabular}
\end{center}
\end{table}

The strongest correlation we find is with the tidal tensor $\hat
T$, which is significant for all three investigated smoothing scales $r$. We also see
strong excess correlation between the galaxy spins and $\left(\hat T\right)^2$ smoothed
with $r = 2\, h^{-1}\mathrm{Mpc}$ and $r = 3\, h^{-1}\mathrm{Mpc}$ and $\hat I$ smoothed
with $r = 4\, h^{-1}\mathrm{Mpc}$. No other excess correlation reaches $4\sigma$ significance.

\subsection{Systematic bias in the faint subsample}

As pointed out in \cite{2009MNRAS.399.1074H}, selection function of a gravitational survey
generally depends on the spatial orientation of a galaxy relative to the observer's line of sight.
Emission from a disk galaxy will suffer more extinction when the galaxy is viewed edge on, potentially
dropping below the survey detection threshold. Because galaxy orientation is sensitive
to the local tidal field, such selection effect can in principle bias the correlations
studied in this work.

\begin{figure}
\center
\includegraphics[width = 0.49 \textwidth]{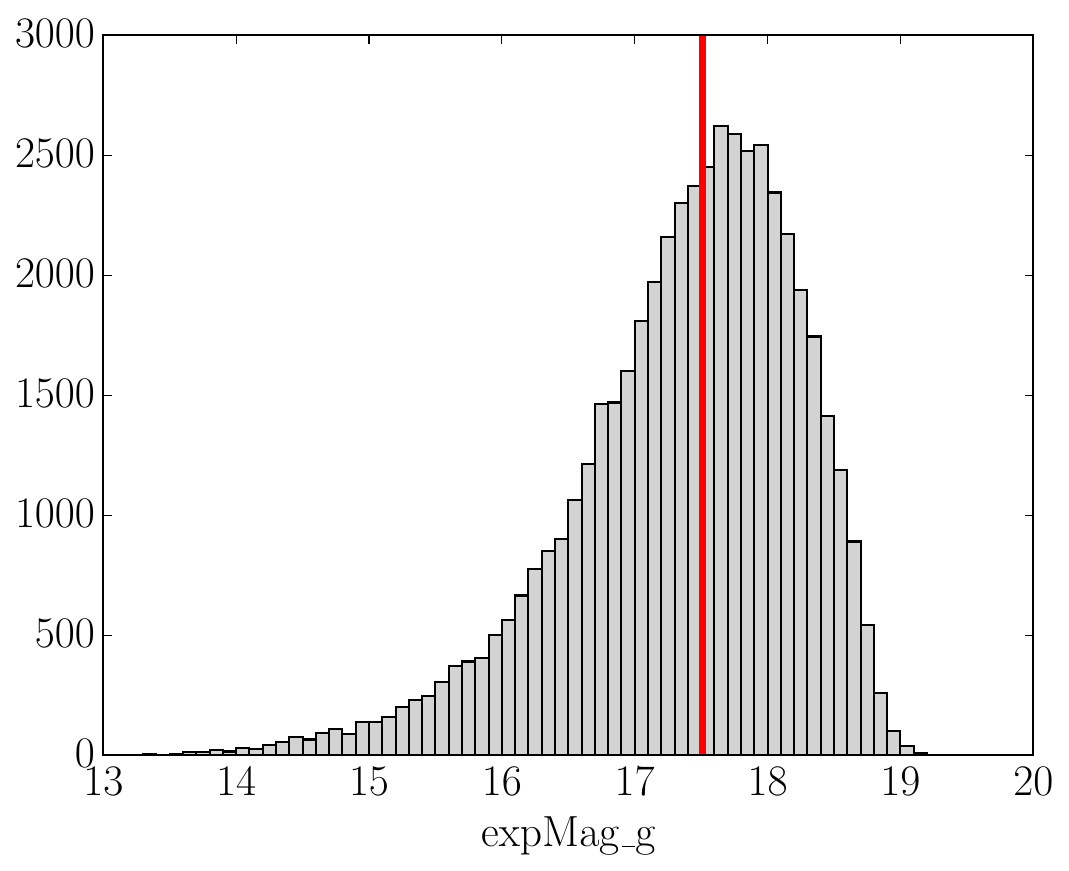}
\caption{
Distribution of the galaxy magnitudes 
\protect\input our.cpt\relax, determined from the exponential fit to the
galaxy images in the green SDSS filter. The red line splits our galaxy sample into the
``bright'' and ``faint'' subsamples of approximately equal sizes.
}
\label{fig:mag_hist}
\end{figure}

To check for presence of this effect in our results, we split our galaxy sample into two
approximately equal sized bins
based on the galaxy magnitude. Specifically, we consider the resulting magnitude from the fit of
the exponential profile to the galaxy image in the green SDSS filter, listed as
\verb|expMag_g| in the SDSS database. 
If the galaxy-orientation-dependent selection effect exists, we expect it to mostly affect
the fainter subsample of galaxies. Comparing the results between the subsamples then
allows us to check for this systematic.
Accidentally, the
splitting magnitude roughly corresponds to the magnitude at which our sample becomes incomplete,
see Fig.~\ref{fig:mag_hist} for the distribution of \verb|expMag_g| in our full galaxy sample.

\begin{figure*}
\center
\includegraphics[width = 0.99 \textwidth]{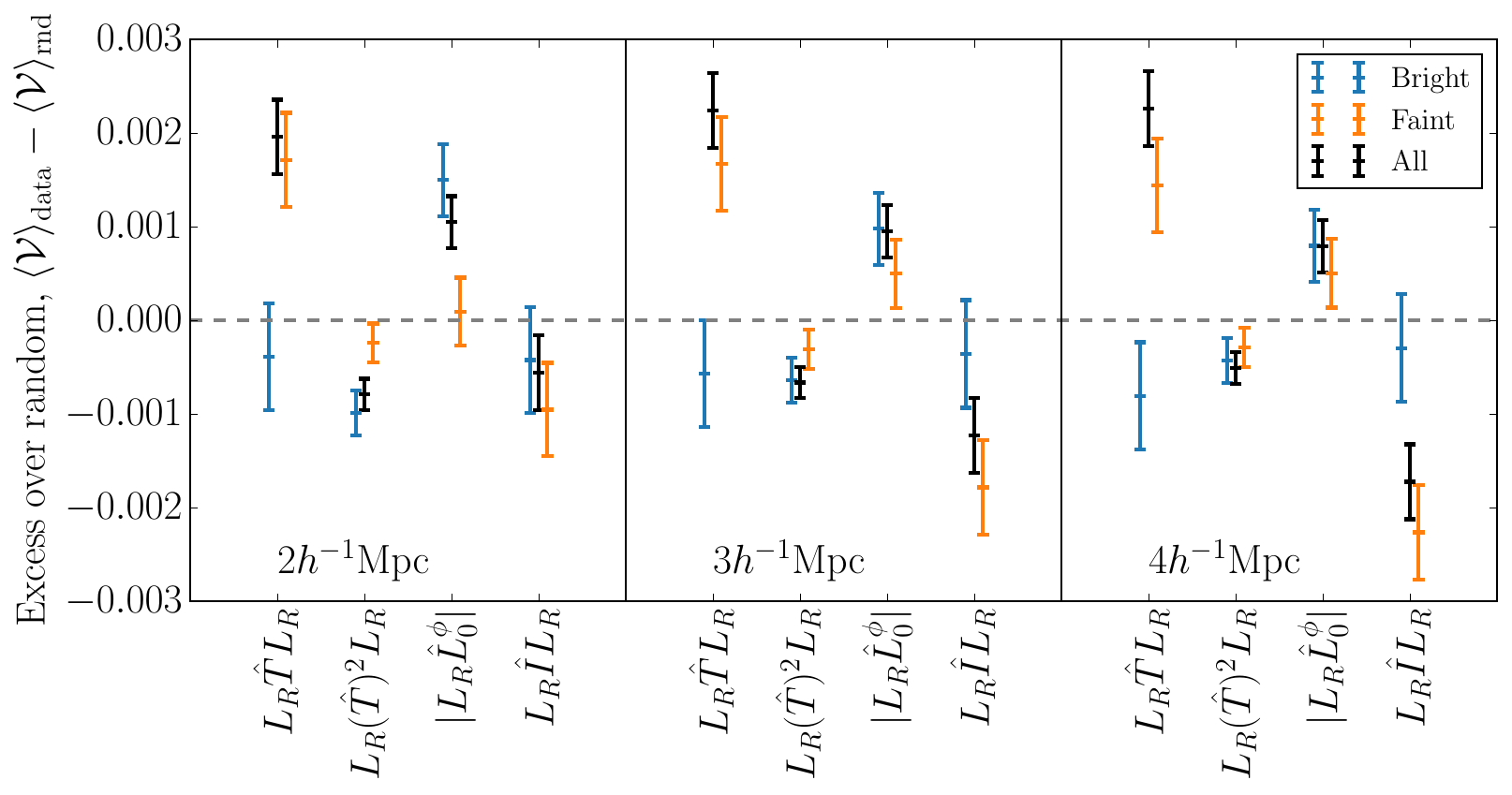}

\caption{
Comparing excess correlations of galaxy spins and initial conditions when using all the galaxies
(black) or only the brighter/fainter (blue/orange) half of galaxies (based on the exponential fit to the
galaxy image in green filter, 
\protect\input our.cpt\relax). For smoothing scales
$2\, h^{-1}\mathrm{Mpc}$ (left),
$3\, h^{-1}\mathrm{Mpc}$ (middle) and
$4\, h^{-1}\mathrm{Mpc}$ (right)
and the four observables with the most significant detection in the full sample.
}
\label{fig:study_mag_splits}
\end{figure*}

In Fig.~\ref{fig:study_mag_splits}
we plot the excess correlation for the four observables with the largest detection
significance in Table~\ref{tab:sigmas} --- 
$L_R\hat TL_R$,
$L_R(\hat T)^2L_R$,
$L_R\hat IL_R$ and
$|L_RL_0^{\phi}|$
--- and all three smoothing scales for both the full
galaxy sample and the bright/faint subsamples.

We find that the detections of excess correlation in
$L_R\hat TL_R$ and
$L_R\hat IL_R$
are clearly driven by the faint galaxy subsample. In the bright subsample, these observables are consistent
with zero excess, which suggests that the excess correlations we found above
arise because of an unaccounted-for systematic effect, such as the 
orientation-dependent selection bias mentioned above.

For 
$L_R(\hat T)^2L_R$ and
$|L_RL_0^{\phi}|$
we find the opposite behavior: the detections are driven by the bright sample, with
the faint sample being generally consistent with no excess. 
As we assume that the bright subsample is mostly complete,
we further focus
our attention only on this half of the galaxies. We further also only consider
$L_R(\hat T)^2L_R$ and
$|L_RL_0^{\phi}|$
constructed from the initial conditions smoothed with $r =
2\,h^{-1}\mathrm{Mpc}$ that leads to the most significant detection. The detection
significances are then formally 4.2$\sigma$ and 4.0$\sigma$, respectively.

We also note that significance of no other observable was notably boosted by restriction to the bright
subsample of the galaxies.

\subsection{Other systematic checks}

\begin{figure*}
\center
\includegraphics[width = 0.89 \textwidth]{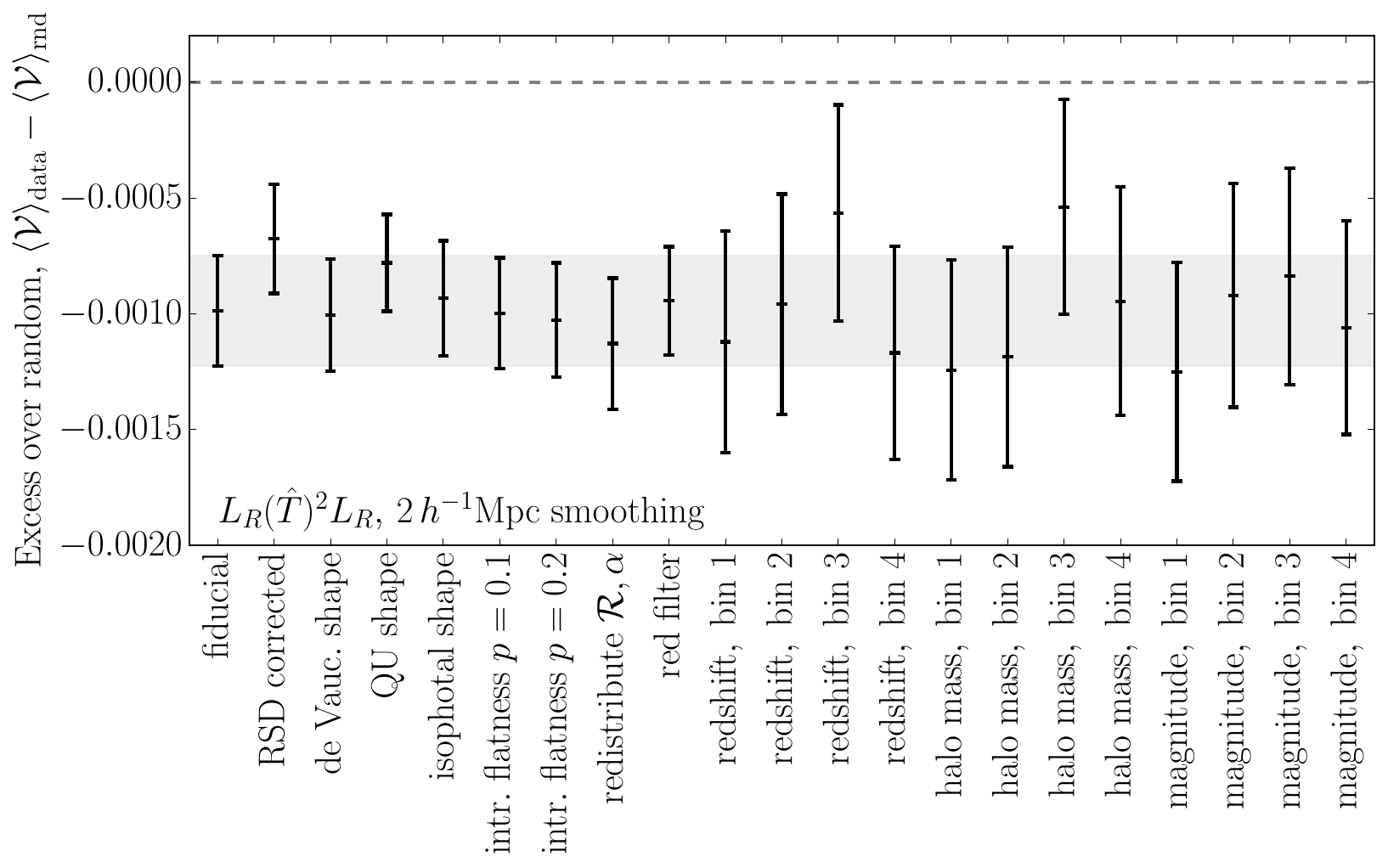}
\includegraphics[width = 0.89 \textwidth]{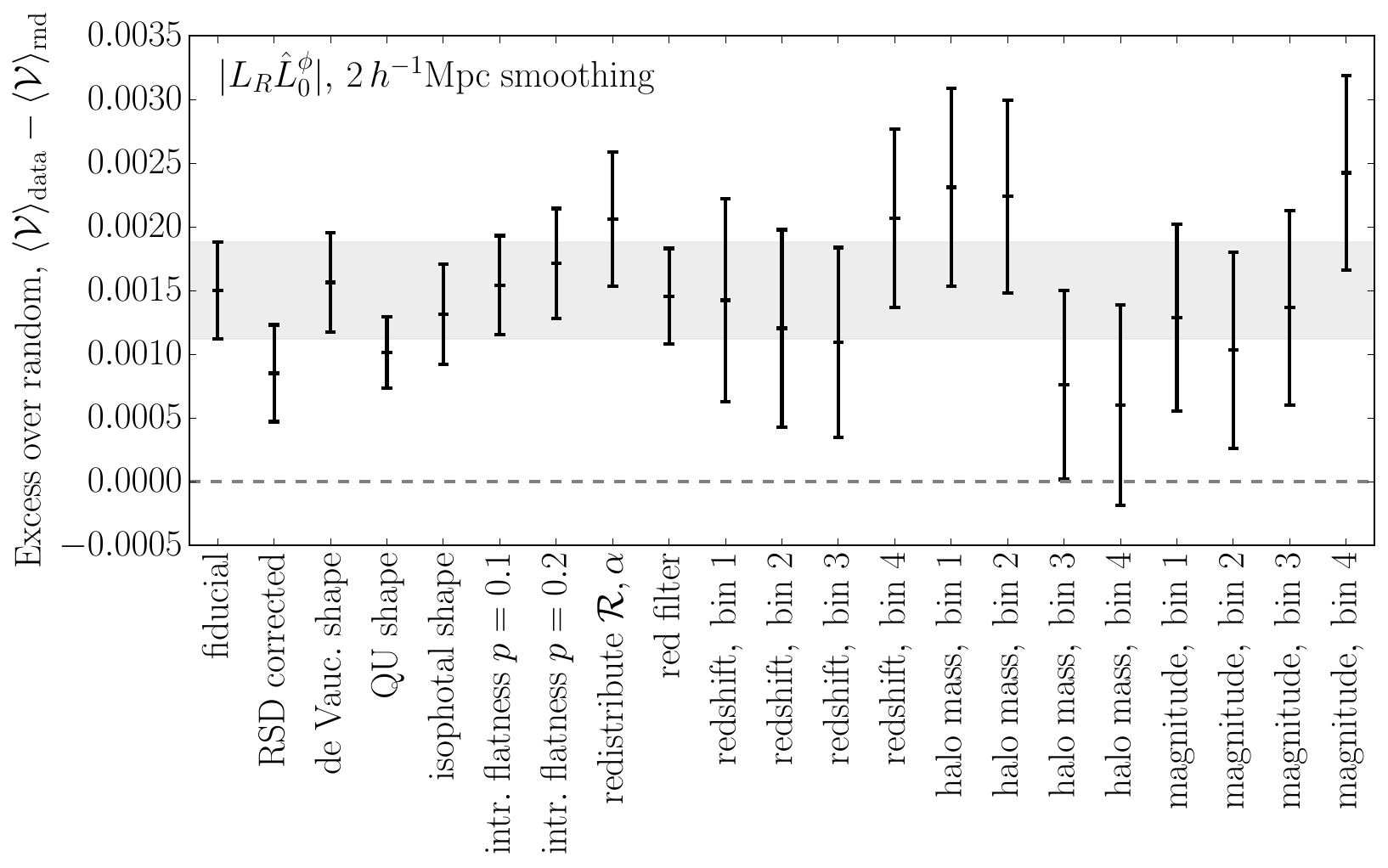}

\caption{
Investigating systematic effects:
Comparing excess correlations for $L_R (\hat T)^2 L_R$ (top) and $\left| L_R \hat L_0^\phi
\right|$ (bottom), both
smoothed with $r = 2\, h^{-1}\mathrm{Mpc}$, for the fiducial analysis (leftmost point, also the gray
band) and various changes to the analysis. We show how the results change when we correct for redshift space
distortions (RSD), when we use other determinations of the galaxy shape (fits of de Vaucouleurs
profile, Stokes parameters $Q$ and $U$ or isophotal ellipticities), when we consider intrinsic flatness with $p$ of 0.1 or
0.2, when we redistribute the axis ratio and position angle to form a uniform
distribution, or when we use red filter images as the baseline. The last twelve points
show results after splitting the data into four bins in either redshift, halo mass
or
magnitude 
\protect\input our.cpt\relax
(bin 1 contains the closest / lightest / brightest galaxies). Dashed line represents no excess
correlation.
}
\label{fig:systematics}
\end{figure*}

While a full study of all possible systematic effects goes beyond the scope of this work,
to check for potential other systematics
we perform several other tests.
As motivated above, we only consider excess correlations of
$L_R(\hat T)^2L_R$ and
$|L_RL_0^{\phi}|$
built from the initial conditions
smoothed with the smoothing scale $2\, h^{-1}\mathrm{Mpc}$. The results are shown in the
top
($L_R(\hat T)^2L_R$)
and 
bottom 
($|L_RL_0^{\phi}|$)
parts of
Fig.~\ref{fig:systematics} and we give more details about the tests in what follows;
we discuss the tests in the order in which they appear in Fig.~\ref{fig:systematics}. The
left-most point in both parts of Fig.~\ref{fig:systematics} represents our result from
the previous section.
In this section we do not consider the galaxies from the ``faint'' subsample at all, because of
the issues uncovered in the previous section.

Up to this point we completely ignored the redshift space distortions (RSD).
If we correct redshift of each galaxy by the same amount the ELUCID Collaboration
corrected the redshift on the nearest dark matter halo, the detection significance 
goes down to 2.9$\sigma$ for 
$L_R(\hat T)^2L_R$
and 2.2$\sigma$ for
$|L_RL_0^{\phi}|$. The shift in significance gives us
an estimate of how big is the effect of correcting the (linear) RSDs, though one has to
keep in mind there are also non-linear
contributions to the RSD that can potentially contribute more (in either direction).

In the next three tests we investigated what happens when we use different determinations of
galaxy shapes. Instead of the results of the exponential fit used as the default, we
consider results of fits assuming the de Vaucouleurs profile of galaxies, 
the Stokes parameters $Q$ and $U$ based on weighted second moments of galaxy intensity
or
ellipticities based on galaxy isophotes (25 magnitudes per square arcsecond).
The latter two determinations of galaxy shapes are based on galaxy images
uncorrected for the point spread function (PSF). For the fits of the de Vaucouleurs
profile we do not see a big change in the results. Using the Stokes parameters decreases
the detection significance by 0.3$\sigma$--0.4$\sigma$. It may appear from
Fig.~\ref{fig:systematics} that
the effect is bigger, but using the Stokes parameters shrinks the error bars due
to a different distribution of $\mathcal{R}$ and this must be taken into account.
Finally, analysis based on the isophotal quantities gives detection significance smaller
by 0.4$\sigma$--0.6$\sigma$.

Next two tests investigated the assumption of galaxies as infinitely flat disks by
assuming constant intrinsic flatness either $p = 0.1$ or $p = 0.2$ and correcting the galaxy
inclinations according to \eqref{intrinsic_flatness_replacement}. We do not find
significant changes. In a similar vein, we tried ``isotropising'' the distribution of
$\alpha$ and $\mathcal{R}$ by uniformly redistributing each in the interval [0, $\pi$]
resp. [0, 1] while keeping their original ordering unchanged \cite{Lee:2000br}. So for
example if the first galaxy had larger $\mathcal{R}$ than the second galaxy before the
redistribution, it will have larger $\mathcal{R}$ after the redistribution.
Again, we do not see our result notably changed.

We also repeated the analysis with galaxy shape parameters determined from the red SDSS images,
to find only minute change to the results

Finally, we further split the bright galaxy subsample into four bins in one of
three different ways: either based
on the galaxy redshift, or on the mass of the corresponding ELUCID halo or finally
on the \verb|expMag_g| magnitude. In all bins we find results consistent with the fiducial
analysis, with the same sign of the excess correlation in all bins. No clear trend is detected and the
scatter between the bins serves as a rough sanity check on the size of our estimated error
bars.

\section{Discussion}
\label{sec:discuss}

In this work we studied correlations between directions of galaxy angular momenta
determined from galaxy shapes (specifically axis ratios and position angles fit from
galaxy images) and various observables
built from the initial conditions (density and gravitational potential), both based on
SDSS data.

When considering the full galaxy sample, we found a formally statistically significant
correlation between the radial component of the galaxy spins $\vec L_R$ and the tidal
field in the vicinity of the protohalo. However, we find that this detection is driven by
the faintest half of our galaxy sample and no significant excess is detected when using only the brightest
half of the galaxies. This suggests that we are in fact detecting a systematic effect,
possibly a galaxy-orientation-dependent selection function \cite{2009MNRAS.399.1074H}.

When we restrict our attention to the brightest half of the galaxies, where we do not
expect this effect to matter, we see about 4$\sigma$ hints of excess correlation in 
$L_R(\hat T)^2L_R$ and
$|L_RL_0^{\phi}|$,
i.e. between the radial components of the galaxy spins $\vec L_R$ and the second power of
the unit traceless tidal tensor, respectively the intermediate principal axis vector of
this tensor. Significance of
these correlations drops bellow 3$\sigma$ when we correct for the linear redshift space
distortions. To test for further systematics, we altered the analysis in various ways and
split the galaxies in bins depending on either galaxy redshift, halo mass or magnitude
\verb|expMag_g|. We did not find any other issues.

One of the checks we performed was using galaxy image axis ratios obtained from galaxy
images uncorrected for the point spread function: either determined from the Stokes
parameters $Q$ and $U$ as weighted second moments of galaxy intensity or from the
isophotal quantities. We find that the detection significance drops by only
0.3$\sigma$--0.4$\sigma$ respectively 0.4$\sigma$--0.6$\sigma$.  While additive PSF
systematics in fits of the de Vaucouleurs profile to SDSS galaxy images are important for
studies of intrinsic alignments \cite{Singh:2015sva}, this result suggests that such
effect is not dominant in our cross-correlation study at current sensitivities.

One must keep in mind these systematic checks are not exhaustive and our results can still
be contaminated by systematics. For example, based on our analysis we suspect that the SDSS
galaxy sample is missing faint, edge-on galaxies. As a consequence, masses of 
galaxy groups in regions where the tidal field orientation leads to
preferentially edge-on galaxies are expected to be underestimated by ELUCID.
It is unclear to what extent such a bias affects the
correlations we study without performing an exhaustive, dedicated study that goes beyond the
scope of this work. To fully understand this and other systematics, it would be necessary
to start with a galaxy simulation that includes galaxy shapes, create a galaxy catalog,
use it to reconstruct the initial conditions similarly to what ELUCID did and only then
perform the analysis of this work.
To name just a few other potential systematics, shape determination from the SDSS galaxy
image can be biased for the edge-on galaxies and the same might in principle be true also for
the redshift determination. Again, full determination of how these effects affect our
results seems to require a separate study.

Despite these worries about additional systematic effects, we point out that the two
correlations our analysis picks out as the most promising are the two correlations
expected from the tidal torque theory arguments \cite{Lee:2000br}. Tidal torque theory
predicts preferred alignment of the galaxy spin with the intermediate axis of the tidal
tensor, which translates into a deficit correlation in $L_R(\hat T)^2L_R$ and excess
correlation in $|L_RL_0^{\phi}|$, which both agree with our results
(Fig.~\ref{fig:study_mag_splits}).

We find that excess signals are mutually compatible when using either $\vec L_R$ or $\vec L_T$,
galaxy spin vectors parallel and tangential to the line of sight. The error bars are
always larger for correlations including $\vec L_T$, which is understandable given that
the position angle $\alpha$ acts as an additional source of uncertainty. We also find that
observables built from the tidal tensor $\hat T$ are more promising than those built from
$\hat I$, the second derivatives of the density field. In hindsight, this is easy to
understand, because from the Poisson equation we see that relative to $\hat T$, the $\hat I$ field puts higher
weight on the small scale data, which are noisier in the reconstruction.

Overall, given SDSS catalog of galaxy shapes and initial conditions reconstructed from
SDSS data,
we are not able to find a statistically significant correlation between the two. 
Beyond the systematics study mentioned in the discussion,
one area of future study is comparing reconstruction of initial
conditions in a simulation, once with and once without galaxy spin data, to study the
potential information gain.
It will also be interesting to see how our analysis improves with the extended galaxy
catalogs from DESI and potentially better reconstruction. 
Our work picks two observables that such studies should focus on.

\acknowledgements{
We thank anonymous referees for their very valuable suggestions, especially regarding various
systematic effects.
We also thank M. Takada and participants of the YITP workshop YITP-T-21-06 on "Galaxy shape
statistics and Cosmology" for useful discussions.

This work was supported by Natural Sciences and Engineering Research
Council of Canada (NSERC) grant CITA 490888. 
P.M. was additionally supported by Vincent and Beatrice Tremaine Fellowship.
U.-L.P. received
additional support from Ontario Research Fund-research Excellence Program grant RE09-024,
NSERC grants RGPIN-2019-067, 523638-201, CRDPJ 523638-2, Canadian Institute for Advanced
Research grants FS21-146 and APPT, Canadian Foundation for Innovation grant IOF-33526,
Simons Foundation grant 568354, Thoth Technology Inc., and the Alexander von Humboldt
Foundation.
H.-R.Y. additionally acknowledges support
from National Natural Science Foundation of China grant 11903021. 
}

\appendix

\bibliography{shape_ic_correlation}

\begin{thebibliography}{54}%
\makeatletter
\providecommand \@ifxundefined [1]{%
 \@ifx{#1\undefined}
}%
\providecommand \@ifnum [1]{%
 \ifnum #1\expandafter \@firstoftwo
 \else \expandafter \@secondoftwo
 \fi
}%
\providecommand \@ifx [1]{%
 \ifx #1\expandafter \@firstoftwo
 \else \expandafter \@secondoftwo
 \fi
}%
\providecommand \natexlab [1]{#1}%
\providecommand \enquote  [1]{``#1''}%
\providecommand \bibnamefont  [1]{#1}%
\providecommand \bibfnamefont [1]{#1}%
\providecommand \citenamefont [1]{#1}%
\providecommand \href@noop [0]{\@secondoftwo}%
\providecommand \href [0]{\begingroup \@sanitize@url \@href}%
\providecommand \@href[1]{\@@startlink{#1}\@@href}%
\providecommand \@@href[1]{\endgroup#1\@@endlink}%
\providecommand \@sanitize@url [0]{\catcode `\\12\catcode `\$12\catcode
  `\&12\catcode `\#12\catcode `\^12\catcode `\_12\catcode `\%12\relax}%
\providecommand \@@startlink[1]{}%
\providecommand \@@endlink[0]{}%
\providecommand \url  [0]{\begingroup\@sanitize@url \@url }%
\providecommand \@url [1]{\endgroup\@href {#1}{\urlprefix }}%
\providecommand \urlprefix  [0]{URL }%
\providecommand \Eprint [0]{\href }%
\providecommand \doibase [0]{http://dx.doi.org/}%
\providecommand \selectlanguage [0]{\@gobble}%
\providecommand \bibinfo  [0]{\@secondoftwo}%
\providecommand \bibfield  [0]{\@secondoftwo}%
\providecommand \translation [1]{[#1]}%
\providecommand \BibitemOpen [0]{}%
\providecommand \bibitemStop [0]{}%
\providecommand \bibitemNoStop [0]{.\EOS\space}%
\providecommand \EOS [0]{\spacefactor3000\relax}%
\providecommand \BibitemShut  [1]{\csname bibitem#1\endcsname}%
\let\auto@bib@innerbib\@empty
%</preamble>
\bibitem [{\citenamefont {Geller}\ and\ \citenamefont
  {Huchra}(1989)}]{Geller:1989da}%
  \BibitemOpen
  \bibfield  {author} {\bibinfo {author} {\bibfnamefont {M.~J.}\ \bibnamefont
  {Geller}}\ and\ \bibinfo {author} {\bibfnamefont {J.~P.}\ \bibnamefont
  {Huchra}},\ }\href {\doibase 10.1126/science.246.4932.897} {\bibfield
  {journal} {\bibinfo  {journal} {Science}\ }\textbf {\bibinfo {volume}
  {246}},\ \bibinfo {pages} {897} (\bibinfo {year} {1989})}\BibitemShut
  {NoStop}%
\bibitem [{\citenamefont {Colless}\ \emph {et~al.}(2003)\citenamefont {Colless}
  \emph {et~al.}}]{Colless:2003wz}%
  \BibitemOpen
  \bibfield  {author} {\bibinfo {author} {\bibfnamefont {M.}~\bibnamefont
  {Colless}} \emph {et~al.},\ }\href@noop {} {\  (\bibinfo {year} {2003})},\
  \Eprint {http://arxiv.org/abs/astro-ph/0306581} {arXiv:astro-ph/0306581}
  \BibitemShut {NoStop}%
\bibitem [{\citenamefont {Driver}\ \emph {et~al.}(2011)\citenamefont {Driver}
  \emph {et~al.}}]{Driver:2010zb}%
  \BibitemOpen
  \bibfield  {author} {\bibinfo {author} {\bibfnamefont {S.~P.}\ \bibnamefont
  {Driver}} \emph {et~al.},\ }\href {\doibase 10.1111/j.1365-2966.2010.18188.x}
  {\bibfield  {journal} {\bibinfo  {journal} {Mon. Not. Roy. Astron. Soc.}\
  }\textbf {\bibinfo {volume} {413}},\ \bibinfo {pages} {971} (\bibinfo {year}
  {2011})},\ \Eprint {http://arxiv.org/abs/1009.0614} {arXiv:1009.0614
  [astro-ph.CO]} \BibitemShut {NoStop}%
\bibitem [{\citenamefont {{York}}\ \emph {et~al.}(2000)\citenamefont {{York}}
  \emph {et~al.}}]{2000AJ....120.1579Y}%
  \BibitemOpen
  \bibfield  {author} {\bibinfo {author} {\bibfnamefont {D.~G.}\ \bibnamefont
  {{York}}} \emph {et~al.} (\bibinfo {collaboration} {SDSS}),\ }\href@noop {}
  {\bibfield  {journal} {\bibinfo  {journal} {Astron. J.}\ }\textbf {\bibinfo
  {volume} {120}},\ \bibinfo {pages} {1579} (\bibinfo {year}
  {2000})}\BibitemShut {NoStop}%
\bibitem [{\citenamefont {Levi}\ \emph {et~al.}(2013)\citenamefont {Levi} \emph
  {et~al.}}]{Levi:2013gra}%
  \BibitemOpen
  \bibfield  {author} {\bibinfo {author} {\bibfnamefont {M.}~\bibnamefont
  {Levi}} \emph {et~al.} (\bibinfo {collaboration} {DESI}),\ }\href@noop {} {\
  (\bibinfo {year} {2013})},\ \Eprint {http://arxiv.org/abs/1308.0847}
  {arXiv:1308.0847 [astro-ph.CO]} \BibitemShut {NoStop}%
%%CITATION = ARXIV:1308.0847;%%
\bibitem [{\citenamefont {Abell}\ \emph {et~al.}(2009)\citenamefont {Abell}
  \emph {et~al.}}]{LSSTScience:2009jmu}%
  \BibitemOpen
  \bibfield  {author} {\bibinfo {author} {\bibfnamefont {P.~A.}\ \bibnamefont
  {Abell}} \emph {et~al.} (\bibinfo {collaboration} {LSST Science, LSST
  Project}),\ }\href@noop {} {\  (\bibinfo {year} {2009})},\ \Eprint
  {http://arxiv.org/abs/0912.0201} {arXiv:0912.0201 [astro-ph.IM]} \BibitemShut
  {NoStop}%
\bibitem [{\citenamefont {Lee}\ and\ \citenamefont {Pen}(2000)}]{Lee:1999ii}%
  \BibitemOpen
  \bibfield  {author} {\bibinfo {author} {\bibfnamefont {J.}~\bibnamefont
  {Lee}}\ and\ \bibinfo {author} {\bibfnamefont {U.-L.}\ \bibnamefont {Pen}},\
  }\href@noop {} {\bibfield  {journal} {\bibinfo  {journal} {Astrophys. J.}\
  }\textbf {\bibinfo {volume} {532}},\ \bibinfo {pages} {L5} (\bibinfo {year}
  {2000})}\BibitemShut {NoStop}%
\bibitem [{\citenamefont {Lee}\ and\ \citenamefont {Pen}(2001)}]{Lee:2000br}%
  \BibitemOpen
  \bibfield  {author} {\bibinfo {author} {\bibfnamefont {J.}~\bibnamefont
  {Lee}}\ and\ \bibinfo {author} {\bibfnamefont {U.-L.}\ \bibnamefont {Pen}},\
  }\href@noop {} {\bibfield  {journal} {\bibinfo  {journal} {Astrophys. J.}\
  }\textbf {\bibinfo {volume} {555}},\ \bibinfo {pages} {106} (\bibinfo {year}
  {2001})}\BibitemShut {NoStop}%
\bibitem [{\citenamefont {Yu}\ \emph {et~al.}(2020)\citenamefont {Yu},
  \citenamefont {Motloch}, \citenamefont {Pen}, \citenamefont {Yu},
  \citenamefont {Wang}, \citenamefont {Mo}, \citenamefont {Yang},\ and\
  \citenamefont {Jing}}]{Yu:2019bsd}%
  \BibitemOpen
  \bibfield  {author} {\bibinfo {author} {\bibfnamefont {H.-R.}\ \bibnamefont
  {Yu}}, \bibinfo {author} {\bibfnamefont {P.}~\bibnamefont {Motloch}},
  \bibinfo {author} {\bibfnamefont {U.-L.}\ \bibnamefont {Pen}}, \bibinfo
  {author} {\bibfnamefont {Y.}~\bibnamefont {Yu}}, \bibinfo {author}
  {\bibfnamefont {H.}~\bibnamefont {Wang}}, \bibinfo {author} {\bibfnamefont
  {H.}~\bibnamefont {Mo}}, \bibinfo {author} {\bibfnamefont {X.}~\bibnamefont
  {Yang}}, \ and\ \bibinfo {author} {\bibfnamefont {Y.}~\bibnamefont {Jing}},\
  }\href@noop {} {\bibfield  {journal} {\bibinfo  {journal} {Phys. Rev. Lett.}\
  }\textbf {\bibinfo {volume} {124}},\ \bibinfo {pages} {101302} (\bibinfo
  {year} {2020})}\BibitemShut {NoStop}%
%%CITATION = ARXIV:1904.01029;%%
\bibitem [{\citenamefont {Yu}\ \emph {et~al.}(2019)\citenamefont {Yu},
  \citenamefont {Pen},\ and\ \citenamefont {Wang}}]{Yu:2018llx}%
  \BibitemOpen
  \bibfield  {author} {\bibinfo {author} {\bibfnamefont {H.-R.}\ \bibnamefont
  {Yu}}, \bibinfo {author} {\bibfnamefont {U.-L.}\ \bibnamefont {Pen}}, \ and\
  \bibinfo {author} {\bibfnamefont {X.}~\bibnamefont {Wang}},\ }\href@noop {}
  {\bibfield  {journal} {\bibinfo  {journal} {Phys. Rev.}\ }\textbf {\bibinfo
  {volume} {D99}},\ \bibinfo {pages} {123532} (\bibinfo {year}
  {2019})}\BibitemShut {NoStop}%
\bibitem [{\citenamefont {Biagetti}\ and\ \citenamefont
  {Orlando}(2020)}]{Biagetti:2020lpx}%
  \BibitemOpen
  \bibfield  {author} {\bibinfo {author} {\bibfnamefont {M.}~\bibnamefont
  {Biagetti}}\ and\ \bibinfo {author} {\bibfnamefont {G.}~\bibnamefont
  {Orlando}},\ }\href@noop {} {\bibfield  {journal} {\bibinfo  {journal} {J.
  Cosmo. Astropart. Phys.}\ }\textbf {\bibinfo {volume} {07}},\ \bibinfo
  {pages} {005} (\bibinfo {year} {2020})}\BibitemShut {NoStop}%
\bibitem [{\citenamefont {Schmidt}\ \emph {et~al.}(2015)\citenamefont
  {Schmidt}, \citenamefont {Chisari},\ and\ \citenamefont
  {Dvorkin}}]{Schmidt:2015xka}%
  \BibitemOpen
  \bibfield  {author} {\bibinfo {author} {\bibfnamefont {F.}~\bibnamefont
  {Schmidt}}, \bibinfo {author} {\bibfnamefont {N.~E.}\ \bibnamefont
  {Chisari}}, \ and\ \bibinfo {author} {\bibfnamefont {C.}~\bibnamefont
  {Dvorkin}},\ }\href@noop {} {\bibfield  {journal} {\bibinfo  {journal} {J.
  Cosmo. Astropart. Phys.}\ }\textbf {\bibinfo {volume} {1510}},\ \bibinfo
  {pages} {032} (\bibinfo {year} {2015})}\BibitemShut {NoStop}%
\bibitem [{\citenamefont {Peebles}(1969)}]{Peebles:1969jm}%
  \BibitemOpen
  \bibfield  {author} {\bibinfo {author} {\bibfnamefont {P.}~\bibnamefont
  {Peebles}},\ }\href@noop {} {\bibfield  {journal} {\bibinfo  {journal}
  {Astrophys. J.}\ }\textbf {\bibinfo {volume} {155}},\ \bibinfo {pages} {393}
  (\bibinfo {year} {1969})}\BibitemShut {NoStop}%
\bibitem [{\citenamefont {{Doroshkevich}}(1970)}]{1970Ap......6..320D}%
  \BibitemOpen
  \bibfield  {author} {\bibinfo {author} {\bibfnamefont {A.~G.}\ \bibnamefont
  {{Doroshkevich}}},\ }\href@noop {} {\bibfield  {journal} {\bibinfo  {journal}
  {Astrophys.}\ }\textbf {\bibinfo {volume} {6}},\ \bibinfo {pages} {320}
  (\bibinfo {year} {1970})}\BibitemShut {NoStop}%
\bibitem [{\citenamefont {{White}}(1984)}]{1984ApJ...286...38W}%
  \BibitemOpen
  \bibfield  {author} {\bibinfo {author} {\bibfnamefont {S.~D.~M.}\
  \bibnamefont {{White}}},\ }\href@noop {} {\bibfield  {journal} {\bibinfo
  {journal} {Astrophys. J.}\ }\textbf {\bibinfo {volume} {286}},\ \bibinfo
  {pages} {38} (\bibinfo {year} {1984})}\BibitemShut {NoStop}%
\bibitem [{\citenamefont {Porciani}\ \emph
  {et~al.}(2002{\natexlab{a}})\citenamefont {Porciani}, \citenamefont {Dekel},\
  and\ \citenamefont {Hoffman}}]{Porciani:2001db}%
  \BibitemOpen
  \bibfield  {author} {\bibinfo {author} {\bibfnamefont {C.}~\bibnamefont
  {Porciani}}, \bibinfo {author} {\bibfnamefont {A.}~\bibnamefont {Dekel}}, \
  and\ \bibinfo {author} {\bibfnamefont {Y.}~\bibnamefont {Hoffman}},\
  }\href@noop {} {\bibfield  {journal} {\bibinfo  {journal} {Mon. Not. R.
  Astron. Soc.}\ }\textbf {\bibinfo {volume} {332}},\ \bibinfo {pages} {325}
  (\bibinfo {year} {2002}{\natexlab{a}})}\BibitemShut {NoStop}%
\bibitem [{\citenamefont {Porciani}\ \emph
  {et~al.}(2002{\natexlab{b}})\citenamefont {Porciani}, \citenamefont {Dekel},\
  and\ \citenamefont {Hoffman}}]{Porciani:2001er}%
  \BibitemOpen
  \bibfield  {author} {\bibinfo {author} {\bibfnamefont {C.}~\bibnamefont
  {Porciani}}, \bibinfo {author} {\bibfnamefont {A.}~\bibnamefont {Dekel}}, \
  and\ \bibinfo {author} {\bibfnamefont {Y.}~\bibnamefont {Hoffman}},\
  }\href@noop {} {\bibfield  {journal} {\bibinfo  {journal} {Mon. Not. R.
  Astron. Soc.}\ }\textbf {\bibinfo {volume} {332}},\ \bibinfo {pages} {339}
  (\bibinfo {year} {2002}{\natexlab{b}})}\BibitemShut {NoStop}%
\bibitem [{\citenamefont {Krolewski}\ \emph {et~al.}(2019)\citenamefont
  {Krolewski} \emph {et~al.}}]{Krolewski:2019bfv}%
  \BibitemOpen
  \bibfield  {author} {\bibinfo {author} {\bibfnamefont {A.}~\bibnamefont
  {Krolewski}} \emph {et~al.},\ }\href@noop {} {\bibfield  {journal} {\bibinfo
  {journal} {Astrophys. J.}\ }\textbf {\bibinfo {volume} {876}},\ \bibinfo
  {pages} {52} (\bibinfo {year} {2019})}\BibitemShut {NoStop}%
\bibitem [{\citenamefont {{DeFelippis}}\ \emph {et~al.}(2017)\citenamefont
  {{DeFelippis}}, \citenamefont {{Genel}}, \citenamefont {{Bryan}},\ and\
  \citenamefont {{Fall}}}]{2017ApJ...841...16D}%
  \BibitemOpen
  \bibfield  {author} {\bibinfo {author} {\bibfnamefont {D.}~\bibnamefont
  {{DeFelippis}}}, \bibinfo {author} {\bibfnamefont {S.}~\bibnamefont
  {{Genel}}}, \bibinfo {author} {\bibfnamefont {G.~L.}\ \bibnamefont
  {{Bryan}}}, \ and\ \bibinfo {author} {\bibfnamefont {S.~M.}\ \bibnamefont
  {{Fall}}},\ }\href {\doibase 10.3847/1538-4357/aa6dfc} {\bibfield  {journal}
  {\bibinfo  {journal} {\apj}\ }\textbf {\bibinfo {volume} {841}},\ \bibinfo
  {eid} {16} (\bibinfo {year} {2017})},\ \Eprint
  {http://arxiv.org/abs/1703.03806} {arXiv:1703.03806 [astro-ph.GA]}
  \BibitemShut {NoStop}%
\bibitem [{\citenamefont {Zhang}\ \emph {et~al.}(2015)\citenamefont {Zhang}
  \emph {et~al.}}]{Zhang:2014rju}%
  \BibitemOpen
  \bibfield  {author} {\bibinfo {author} {\bibfnamefont {Y.}~\bibnamefont
  {Zhang}} \emph {et~al.},\ }\href@noop {} {\bibfield  {journal} {\bibinfo
  {journal} {Astrophys. J.}\ }\textbf {\bibinfo {volume} {798}},\ \bibinfo
  {pages} {17} (\bibinfo {year} {2015})}\BibitemShut {NoStop}%
\bibitem [{\citenamefont {Veena}\ \emph {et~al.}(2018)\citenamefont {Veena},
  \citenamefont {Cautun}, \citenamefont {van~de Weygaert}, \citenamefont
  {Tempel}, \citenamefont {Jones}, \citenamefont {Rieder},\ and\ \citenamefont
  {Frenk}}]{Veena:2018ooo}%
  \BibitemOpen
  \bibfield  {author} {\bibinfo {author} {\bibfnamefont {P.~G.}\ \bibnamefont
  {Veena}}, \bibinfo {author} {\bibfnamefont {M.}~\bibnamefont {Cautun}},
  \bibinfo {author} {\bibfnamefont {R.}~\bibnamefont {van~de Weygaert}},
  \bibinfo {author} {\bibfnamefont {E.}~\bibnamefont {Tempel}}, \bibinfo
  {author} {\bibfnamefont {B.~J.}\ \bibnamefont {Jones}}, \bibinfo {author}
  {\bibfnamefont {S.}~\bibnamefont {Rieder}}, \ and\ \bibinfo {author}
  {\bibfnamefont {C.~S.}\ \bibnamefont {Frenk}},\ }\href@noop {} {\bibfield
  {journal} {\bibinfo  {journal} {Mon. Not. R. Astron. Soc.}\ }\textbf
  {\bibinfo {volume} {481}},\ \bibinfo {pages} {414} (\bibinfo {year}
  {2018})}\BibitemShut {NoStop}%
\bibitem [{\citenamefont {Ganeshaiah~Veena}\ \emph {et~al.}(2019)\citenamefont
  {Ganeshaiah~Veena}, \citenamefont {Cautun}, \citenamefont {Tempel},
  \citenamefont {van~de Weygaert},\ and\ \citenamefont
  {Frenk}}]{Veena:2019ozd}%
  \BibitemOpen
  \bibfield  {author} {\bibinfo {author} {\bibfnamefont {P.}~\bibnamefont
  {Ganeshaiah~Veena}}, \bibinfo {author} {\bibfnamefont {M.}~\bibnamefont
  {Cautun}}, \bibinfo {author} {\bibfnamefont {E.}~\bibnamefont {Tempel}},
  \bibinfo {author} {\bibfnamefont {R.}~\bibnamefont {van~de Weygaert}}, \ and\
  \bibinfo {author} {\bibfnamefont {C.~S.}\ \bibnamefont {Frenk}},\ }\href@noop
  {} {\bibfield  {journal} {\bibinfo  {journal} {Mon. Not. R. Astron. Soc.}\
  }\textbf {\bibinfo {volume} {487}},\ \bibinfo {pages} {1607} (\bibinfo {year}
  {2019})}\BibitemShut {NoStop}%
\bibitem [{\citenamefont {Kraljic}\ \emph {et~al.}(2021)\citenamefont
  {Kraljic}, \citenamefont {Duckworth}, \citenamefont {Tojeiro}, \citenamefont
  {Alam}, \citenamefont {Bizyaev}, \citenamefont {Weijmans}, \citenamefont
  {Boardman},\ and\ \citenamefont {Lane}}]{Kraljic:2021oeg}%
  \BibitemOpen
  \bibfield  {author} {\bibinfo {author} {\bibfnamefont {K.}~\bibnamefont
  {Kraljic}}, \bibinfo {author} {\bibfnamefont {C.}~\bibnamefont {Duckworth}},
  \bibinfo {author} {\bibfnamefont {R.}~\bibnamefont {Tojeiro}}, \bibinfo
  {author} {\bibfnamefont {S.}~\bibnamefont {Alam}}, \bibinfo {author}
  {\bibfnamefont {D.}~\bibnamefont {Bizyaev}}, \bibinfo {author} {\bibfnamefont
  {A.-M.}\ \bibnamefont {Weijmans}}, \bibinfo {author} {\bibfnamefont {N.~F.}\
  \bibnamefont {Boardman}}, \ and\ \bibinfo {author} {\bibfnamefont {R.~R.}\
  \bibnamefont {Lane}},\ }\href {\doibase 10.1093/mnras/stab1109} {\bibfield
  {journal} {\bibinfo  {journal} {Mon. Not. Roy. Astron. Soc.}\ }\textbf
  {\bibinfo {volume} {504}},\ \bibinfo {pages} {4626} (\bibinfo {year}
  {2021})},\ \Eprint {http://arxiv.org/abs/2104.08275} {arXiv:2104.08275
  [astro-ph.GA]} \BibitemShut {NoStop}%
\bibitem [{\citenamefont {Kraljic}\ \emph {et~al.}(2020)\citenamefont
  {Kraljic}, \citenamefont {Dave},\ and\ \citenamefont
  {Pichon}}]{Kraljic:2019lca}%
  \BibitemOpen
  \bibfield  {author} {\bibinfo {author} {\bibfnamefont {K.}~\bibnamefont
  {Kraljic}}, \bibinfo {author} {\bibfnamefont {R.}~\bibnamefont {Dave}}, \
  and\ \bibinfo {author} {\bibfnamefont {C.}~\bibnamefont {Pichon}},\ }\href
  {\doibase 10.1093/mnras/staa250} {\bibfield  {journal} {\bibinfo  {journal}
  {Mon. Not. Roy. Astron. Soc.}\ }\textbf {\bibinfo {volume} {493}},\ \bibinfo
  {pages} {362} (\bibinfo {year} {2020})},\ \Eprint
  {http://arxiv.org/abs/1906.01623} {arXiv:1906.01623 [astro-ph.GA]}
  \BibitemShut {NoStop}%
\bibitem [{\citenamefont {Wang}\ and\ \citenamefont
  {Kang}(2018)}]{Wang:2017tsr}%
  \BibitemOpen
  \bibfield  {author} {\bibinfo {author} {\bibfnamefont {P.}~\bibnamefont
  {Wang}}\ and\ \bibinfo {author} {\bibfnamefont {X.}~\bibnamefont {Kang}},\
  }\href {\doibase 10.1093/mnras/stx2466} {\bibfield  {journal} {\bibinfo
  {journal} {Mon. Not. Roy. Astron. Soc.}\ }\textbf {\bibinfo {volume} {473}},\
  \bibinfo {pages} {1562} (\bibinfo {year} {2018})},\ \Eprint
  {http://arxiv.org/abs/1709.07881} {arXiv:1709.07881 [astro-ph.CO]}
  \BibitemShut {NoStop}%
\bibitem [{\citenamefont {Neyrinck}\ \emph {et~al.}(2020)\citenamefont
  {Neyrinck}, \citenamefont {Aragon-Calvo}, \citenamefont {Falck},
  \citenamefont {Szalay},\ and\ \citenamefont {Wang}}]{Neyrinck:2019uvc}%
  \BibitemOpen
  \bibfield  {author} {\bibinfo {author} {\bibfnamefont {M.~C.}\ \bibnamefont
  {Neyrinck}}, \bibinfo {author} {\bibfnamefont {M.~A.}\ \bibnamefont
  {Aragon-Calvo}}, \bibinfo {author} {\bibfnamefont {B.}~\bibnamefont {Falck}},
  \bibinfo {author} {\bibfnamefont {A.~S.}\ \bibnamefont {Szalay}}, \ and\
  \bibinfo {author} {\bibfnamefont {J.}~\bibnamefont {Wang}},\ }\href {\doibase
  10.21105/astro.1904.03201} {\bibfield  {journal} {\bibinfo  {journal} {Open
  J. Astrophys.}\ }\textbf {\bibinfo {volume} {3}},\ \bibinfo {pages} {3}
  (\bibinfo {year} {2020})},\ \Eprint {http://arxiv.org/abs/1904.03201}
  {arXiv:1904.03201 [astro-ph.CO]} \BibitemShut {NoStop}%
\bibitem [{\citenamefont {Schaefer}(2009)}]{Schaefer:2008xd}%
  \BibitemOpen
  \bibfield  {author} {\bibinfo {author} {\bibfnamefont {B.~M.}\ \bibnamefont
  {Schaefer}},\ }\href@noop {} {\bibfield  {journal} {\bibinfo  {journal} {Int.
  J. Mod. Phys. D}\ }\textbf {\bibinfo {volume} {18}},\ \bibinfo {pages} {173}
  (\bibinfo {year} {2009})}\BibitemShut {NoStop}%
\bibitem [{\citenamefont {Codis}\ \emph {et~al.}(2012)\citenamefont {Codis},
  \citenamefont {Pichon}, \citenamefont {Devriendt}, \citenamefont {Slyz},
  \citenamefont {Pogosyan}, \citenamefont {Dubois},\ and\ \citenamefont
  {Sousbie}}]{Codis:2012ep}%
  \BibitemOpen
  \bibfield  {author} {\bibinfo {author} {\bibfnamefont {S.}~\bibnamefont
  {Codis}}, \bibinfo {author} {\bibfnamefont {C.}~\bibnamefont {Pichon}},
  \bibinfo {author} {\bibfnamefont {J.}~\bibnamefont {Devriendt}}, \bibinfo
  {author} {\bibfnamefont {A.}~\bibnamefont {Slyz}}, \bibinfo {author}
  {\bibfnamefont {D.}~\bibnamefont {Pogosyan}}, \bibinfo {author}
  {\bibfnamefont {Y.}~\bibnamefont {Dubois}}, \ and\ \bibinfo {author}
  {\bibfnamefont {T.}~\bibnamefont {Sousbie}},\ }\href@noop {} {\bibfield
  {journal} {\bibinfo  {journal} {Mon. Not. R. Astron. Soc.}\ }\textbf
  {\bibinfo {volume} {427}},\ \bibinfo {pages} {3320} (\bibinfo {year}
  {2012})}\BibitemShut {NoStop}%
\bibitem [{\citenamefont {Codis}\ \emph {et~al.}(2015)\citenamefont {Codis},
  \citenamefont {Pichon},\ and\ \citenamefont {Pogosyan}}]{Codis:2015tla}%
  \BibitemOpen
  \bibfield  {author} {\bibinfo {author} {\bibfnamefont {S.}~\bibnamefont
  {Codis}}, \bibinfo {author} {\bibfnamefont {C.}~\bibnamefont {Pichon}}, \
  and\ \bibinfo {author} {\bibfnamefont {D.}~\bibnamefont {Pogosyan}},\
  }\href@noop {} {\bibfield  {journal} {\bibinfo  {journal} {Mon. Not. R.
  Astron. Soc.}\ }\textbf {\bibinfo {volume} {452}},\ \bibinfo {pages} {3369}
  (\bibinfo {year} {2015})}\BibitemShut {NoStop}%
\bibitem [{\citenamefont {Welker}\ \emph {et~al.}(2020)\citenamefont {Welker}
  \emph {et~al.}}]{Welker:2019puz}%
  \BibitemOpen
  \bibfield  {author} {\bibinfo {author} {\bibfnamefont {C.}~\bibnamefont
  {Welker}} \emph {et~al.},\ }\href@noop {} {\bibfield  {journal} {\bibinfo
  {journal} {Mon. Not. R. Astron. Soc.}\ }\textbf {\bibinfo {volume} {491}},\
  \bibinfo {pages} {2864} (\bibinfo {year} {2020})}\BibitemShut {NoStop}%
\bibitem [{\citenamefont {Jones}\ \emph {et~al.}(2010)\citenamefont {Jones},
  \citenamefont {van~de Weygaert},\ and\ \citenamefont
  {Aragon-Calvo}}]{Jones:2010cs}%
  \BibitemOpen
  \bibfield  {author} {\bibinfo {author} {\bibfnamefont {B.~J.}\ \bibnamefont
  {Jones}}, \bibinfo {author} {\bibfnamefont {R.}~\bibnamefont {van~de
  Weygaert}}, \ and\ \bibinfo {author} {\bibfnamefont {M.~A.}\ \bibnamefont
  {Aragon-Calvo}},\ }\href@noop {} {\bibfield  {journal} {\bibinfo  {journal}
  {Mon. Not. R. Astron. Soc.}\ }\textbf {\bibinfo {volume} {408}},\ \bibinfo
  {pages} {897} (\bibinfo {year} {2010})}\BibitemShut {NoStop}%
\bibitem [{\citenamefont {Aragon-Calvo}\ \emph {et~al.}(2007)\citenamefont
  {Aragon-Calvo}, \citenamefont {van~de Weygaert}, \citenamefont {Jones},\ and\
  \citenamefont {van~der Hulst}}]{AragonCalvo:2006ay}%
  \BibitemOpen
  \bibfield  {author} {\bibinfo {author} {\bibfnamefont {M.~A.}\ \bibnamefont
  {Aragon-Calvo}}, \bibinfo {author} {\bibfnamefont {R.}~\bibnamefont {van~de
  Weygaert}}, \bibinfo {author} {\bibfnamefont {B.~J.}\ \bibnamefont {Jones}},
  \ and\ \bibinfo {author} {\bibfnamefont {J.}~\bibnamefont {van~der Hulst}},\
  }\href@noop {} {\bibfield  {journal} {\bibinfo  {journal} {Astrophys. J.
  Lett.}\ }\textbf {\bibinfo {volume} {655}},\ \bibinfo {pages} {L5} (\bibinfo
  {year} {2007})}\BibitemShut {NoStop}%
\bibitem [{\citenamefont {Hahn}\ \emph {et~al.}(2007)\citenamefont {Hahn},
  \citenamefont {Carollo}, \citenamefont {Porciani},\ and\ \citenamefont
  {Dekel}}]{Hahn:2007ui}%
  \BibitemOpen
  \bibfield  {author} {\bibinfo {author} {\bibfnamefont {O.}~\bibnamefont
  {Hahn}}, \bibinfo {author} {\bibfnamefont {C.}~\bibnamefont {Carollo}},
  \bibinfo {author} {\bibfnamefont {C.}~\bibnamefont {Porciani}}, \ and\
  \bibinfo {author} {\bibfnamefont {A.}~\bibnamefont {Dekel}},\ }\href@noop {}
  {\bibfield  {journal} {\bibinfo  {journal} {Mon. Not. R. Astron. Soc.}\
  }\textbf {\bibinfo {volume} {381}},\ \bibinfo {pages} {41} (\bibinfo {year}
  {2007})}\BibitemShut {NoStop}%
\bibitem [{\citenamefont {Bett}\ and\ \citenamefont
  {Frenk}(2012)}]{Bett:2011rs}%
  \BibitemOpen
  \bibfield  {author} {\bibinfo {author} {\bibfnamefont {P.~E.}\ \bibnamefont
  {Bett}}\ and\ \bibinfo {author} {\bibfnamefont {C.~S.}\ \bibnamefont
  {Frenk}},\ }\href@noop {} {\bibfield  {journal} {\bibinfo  {journal} {Mon.
  Not. R. Astron. Soc.}\ }\textbf {\bibinfo {volume} {420}},\ \bibinfo {pages}
  {3324} (\bibinfo {year} {2012})}\BibitemShut {NoStop}%
\bibitem [{\citenamefont {Bett}\ and\ \citenamefont
  {Frenk}(2016)}]{Bett:2015aoa}%
  \BibitemOpen
  \bibfield  {author} {\bibinfo {author} {\bibfnamefont {P.~E.}\ \bibnamefont
  {Bett}}\ and\ \bibinfo {author} {\bibfnamefont {C.~S.}\ \bibnamefont
  {Frenk}},\ }\href@noop {} {\bibfield  {journal} {\bibinfo  {journal} {Mon.
  Not. R. Astron. Soc.}\ }\textbf {\bibinfo {volume} {461}},\ \bibinfo {pages}
  {1338} (\bibinfo {year} {2016})}\BibitemShut {NoStop}%
\bibitem [{\citenamefont {Tempel}\ and\ \citenamefont
  {Libeskind}(2013)}]{Tempel:2013gqa}%
  \BibitemOpen
  \bibfield  {author} {\bibinfo {author} {\bibfnamefont {E.}~\bibnamefont
  {Tempel}}\ and\ \bibinfo {author} {\bibfnamefont {N.~I.}\ \bibnamefont
  {Libeskind}},\ }\href@noop {} {\bibfield  {journal} {\bibinfo  {journal}
  {Astrophys. J. Lett.}\ }\textbf {\bibinfo {volume} {775}},\ \bibinfo {pages}
  {L42} (\bibinfo {year} {2013})}\BibitemShut {NoStop}%
\bibitem [{\citenamefont {Wang}\ \emph {et~al.}(2018)\citenamefont {Wang},
  \citenamefont {Guo}, \citenamefont {Kang},\ and\ \citenamefont
  {Libeskind}}]{Wang:2018rlf}%
  \BibitemOpen
  \bibfield  {author} {\bibinfo {author} {\bibfnamefont {P.}~\bibnamefont
  {Wang}}, \bibinfo {author} {\bibfnamefont {Q.}~\bibnamefont {Guo}}, \bibinfo
  {author} {\bibfnamefont {X.}~\bibnamefont {Kang}}, \ and\ \bibinfo {author}
  {\bibfnamefont {N.~I.}\ \bibnamefont {Libeskind}},\ }\href@noop {} {\bibfield
   {journal} {\bibinfo  {journal} {Astrophys. J.}\ }\textbf {\bibinfo {volume}
  {866}},\ \bibinfo {pages} {138} (\bibinfo {year} {2018})}\BibitemShut
  {NoStop}%
\bibitem [{\citenamefont {{Teklu}}\ \emph {et~al.}(2015)\citenamefont {{Teklu}}
  \emph {et~al.}}]{2015ApJ...812...29T}%
  \BibitemOpen
  \bibfield  {author} {\bibinfo {author} {\bibfnamefont {A.~F.}\ \bibnamefont
  {{Teklu}}} \emph {et~al.},\ }\href@noop {} {\bibfield  {journal} {\bibinfo
  {journal} {Astrophys. J.}\ }\textbf {\bibinfo {volume} {812}},\ \bibinfo
  {eid} {29} (\bibinfo {year} {2015})}\BibitemShut {NoStop}%
\bibitem [{\citenamefont {Jiang}\ \emph {et~al.}(2019)\citenamefont {Jiang}
  \emph {et~al.}}]{Jiang:2018ioo}%
  \BibitemOpen
  \bibfield  {author} {\bibinfo {author} {\bibfnamefont {F.}~\bibnamefont
  {Jiang}} \emph {et~al.},\ }\href@noop {} {\bibfield  {journal} {\bibinfo
  {journal} {Mon. Not. R. Astron. Soc.}\ }\textbf {\bibinfo {volume} {488}},\
  \bibinfo {pages} {4801} (\bibinfo {year} {2019})}\BibitemShut {NoStop}%
\bibitem [{\citenamefont {Wu}\ \emph {et~al.}(2021)\citenamefont {Wu},
  \citenamefont {Yu}, \citenamefont {Liao},\ and\ \citenamefont
  {Du}}]{Wu:2020btz}%
  \BibitemOpen
  \bibfield  {author} {\bibinfo {author} {\bibfnamefont {Q.}~\bibnamefont
  {Wu}}, \bibinfo {author} {\bibfnamefont {H.-R.}\ \bibnamefont {Yu}}, \bibinfo
  {author} {\bibfnamefont {S.}~\bibnamefont {Liao}}, \ and\ \bibinfo {author}
  {\bibfnamefont {M.}~\bibnamefont {Du}},\ }\href {\doibase
  10.1103/PhysRevD.103.063522} {\bibfield  {journal} {\bibinfo  {journal}
  {Phys. Rev. D}\ }\textbf {\bibinfo {volume} {103}},\ \bibinfo {pages}
  {063522} (\bibinfo {year} {2021})},\ \Eprint
  {http://arxiv.org/abs/2011.03893} {arXiv:2011.03893 [astro-ph.CO]}
  \BibitemShut {NoStop}%
\bibitem [{\citenamefont {Motloch}\ \emph {et~al.}(2021)\citenamefont
  {Motloch}, \citenamefont {Yu}, \citenamefont {Pen},\ and\ \citenamefont
  {Xie}}]{Motloch:2020qvx}%
  \BibitemOpen
  \bibfield  {author} {\bibinfo {author} {\bibfnamefont {P.}~\bibnamefont
  {Motloch}}, \bibinfo {author} {\bibfnamefont {H.-R.}\ \bibnamefont {Yu}},
  \bibinfo {author} {\bibfnamefont {U.-L.}\ \bibnamefont {Pen}}, \ and\
  \bibinfo {author} {\bibfnamefont {Y.}~\bibnamefont {Xie}},\ }\href {\doibase
  10.1038/s41550-020-01262-3} {\bibfield  {journal} {\bibinfo  {journal}
  {Nature Astron.}\ }\textbf {\bibinfo {volume} {5}},\ \bibinfo {pages} {283}
  (\bibinfo {year} {2021})},\ \Eprint {http://arxiv.org/abs/2003.04800}
  {arXiv:2003.04800 [astro-ph.CO]} \BibitemShut {NoStop}%
\bibitem [{\citenamefont {Wang}\ \emph {et~al.}(2016)\citenamefont {Wang} \emph
  {et~al.}}]{Wang:2016qbz}%
  \BibitemOpen
  \bibfield  {author} {\bibinfo {author} {\bibfnamefont {H.}~\bibnamefont
  {Wang}} \emph {et~al.},\ }\href@noop {} {\bibfield  {journal} {\bibinfo
  {journal} {Astrophys. J.}\ }\textbf {\bibinfo {volume} {831}},\ \bibinfo
  {pages} {164} (\bibinfo {year} {2016})}\BibitemShut {NoStop}%
\bibitem [{\citenamefont {Lee}\ and\ \citenamefont
  {Erdogdu}(2007)}]{Lee:2007jx}%
  \BibitemOpen
  \bibfield  {author} {\bibinfo {author} {\bibfnamefont {J.}~\bibnamefont
  {Lee}}\ and\ \bibinfo {author} {\bibfnamefont {P.}~\bibnamefont {Erdogdu}},\
  }\href@noop {} {\bibfield  {journal} {\bibinfo  {journal} {Astrophys. J.}\
  }\textbf {\bibinfo {volume} {671}},\ \bibinfo {pages} {1248} (\bibinfo {year}
  {2007})}\BibitemShut {NoStop}%
\bibitem [{\citenamefont {Aguado}\ \emph {et~al.}(2019)\citenamefont {Aguado}
  \emph {et~al.}}]{Aguado:2018ynx}%
  \BibitemOpen
  \bibfield  {author} {\bibinfo {author} {\bibfnamefont {D.~S.}\ \bibnamefont
  {Aguado}} \emph {et~al.},\ }\href@noop {} {\bibfield  {journal} {\bibinfo
  {journal} {Astrophys. J. Suppl.}\ }\textbf {\bibinfo {volume} {240}},\
  \bibinfo {pages} {23} (\bibinfo {year} {2019})}\BibitemShut {NoStop}%
\bibitem [{\citenamefont {{Scott}}\ \emph {et~al.}(2018)\citenamefont {{Scott}}
  \emph {et~al.}}]{2018MNRAS.481.2299S}%
  \BibitemOpen
  \bibfield  {author} {\bibinfo {author} {\bibfnamefont {N.}~\bibnamefont
  {{Scott}}} \emph {et~al.},\ }\href@noop {} {\bibfield  {journal} {\bibinfo
  {journal} {Mon. Not. R. Astron. Soc.}\ }\textbf {\bibinfo {volume} {481}},\
  \bibinfo {pages} {2299} (\bibinfo {year} {2018})}\BibitemShut {NoStop}%
\bibitem [{\citenamefont {Pen}\ \emph {et~al.}(2000)\citenamefont {Pen},
  \citenamefont {Lee},\ and\ \citenamefont {Seljak}}]{Pen:2000ug}%
  \BibitemOpen
  \bibfield  {author} {\bibinfo {author} {\bibfnamefont {U.-L.}\ \bibnamefont
  {Pen}}, \bibinfo {author} {\bibfnamefont {J.}~\bibnamefont {Lee}}, \ and\
  \bibinfo {author} {\bibfnamefont {U.}~\bibnamefont {Seljak}},\ }\href
  {\doibase 10.1086/317273} {\bibfield  {journal} {\bibinfo  {journal}
  {Astrophys. J. Lett.}\ }\textbf {\bibinfo {volume} {543}},\ \bibinfo {pages}
  {L107} (\bibinfo {year} {2000})},\ \Eprint
  {http://arxiv.org/abs/astro-ph/0006118} {arXiv:astro-ph/0006118} \BibitemShut
  {NoStop}%
\bibitem [{\citenamefont {{Haynes}}\ and\ \citenamefont
  {{Giovanelli}}(1984)}]{1984AJ.....89..758H}%
  \BibitemOpen
  \bibfield  {author} {\bibinfo {author} {\bibfnamefont {M.~P.}\ \bibnamefont
  {{Haynes}}}\ and\ \bibinfo {author} {\bibfnamefont {R.}~\bibnamefont
  {{Giovanelli}}},\ }\href {\doibase 10.1086/113573} {\bibfield  {journal}
  {\bibinfo  {journal} {Astronomical Journal}\ }\textbf {\bibinfo {volume}
  {89}},\ \bibinfo {pages} {758} (\bibinfo {year} {1984})}\BibitemShut
  {NoStop}%
\bibitem [{\citenamefont {Friedrich}\ \emph {et~al.}(2016)\citenamefont
  {Friedrich}, \citenamefont {Seitz}, \citenamefont {Eifler},\ and\
  \citenamefont {Gruen}}]{Friedrich:2015nga}%
  \BibitemOpen
  \bibfield  {author} {\bibinfo {author} {\bibfnamefont {O.}~\bibnamefont
  {Friedrich}}, \bibinfo {author} {\bibfnamefont {S.}~\bibnamefont {Seitz}},
  \bibinfo {author} {\bibfnamefont {T.~F.}\ \bibnamefont {Eifler}}, \ and\
  \bibinfo {author} {\bibfnamefont {D.}~\bibnamefont {Gruen}},\ }\href
  {\doibase 10.1093/mnras/stv2833} {\bibfield  {journal} {\bibinfo  {journal}
  {Mon. Not. Roy. Astron. Soc.}\ }\textbf {\bibinfo {volume} {456}},\ \bibinfo
  {pages} {2662} (\bibinfo {year} {2016})},\ \Eprint
  {http://arxiv.org/abs/1508.00895} {arXiv:1508.00895 [astro-ph.CO]}
  \BibitemShut {NoStop}%
\bibitem [{\citenamefont {Hui}\ and\ \citenamefont {Zhang}(2002)}]{Hui:2002xt}%
  \BibitemOpen
  \bibfield  {author} {\bibinfo {author} {\bibfnamefont {L.}~\bibnamefont
  {Hui}}\ and\ \bibinfo {author} {\bibfnamefont {J.}~\bibnamefont {Zhang}},\
  }\href@noop {} {\  (\bibinfo {year} {2002})},\ \Eprint
  {http://arxiv.org/abs/astro-ph/0205512} {arXiv:astro-ph/0205512} \BibitemShut
  {NoStop}%
\bibitem [{\citenamefont {{Wang}}\ \emph {et~al.}(2014)\citenamefont {{Wang}},
  \citenamefont {{Mo}}, \citenamefont {{Yang}}, \citenamefont {{Jing}},\ and\
  \citenamefont {{Lin}}}]{2014ApJ...794...94W}%
  \BibitemOpen
  \bibfield  {author} {\bibinfo {author} {\bibfnamefont {H.}~\bibnamefont
  {{Wang}}}, \bibinfo {author} {\bibfnamefont {H.~J.}\ \bibnamefont {{Mo}}},
  \bibinfo {author} {\bibfnamefont {X.}~\bibnamefont {{Yang}}}, \bibinfo
  {author} {\bibfnamefont {Y.~P.}\ \bibnamefont {{Jing}}}, \ and\ \bibinfo
  {author} {\bibfnamefont {W.~P.}\ \bibnamefont {{Lin}}},\ }\href@noop {}
  {\bibfield  {journal} {\bibinfo  {journal} {Astrophys. J.}\ }\textbf
  {\bibinfo {volume} {794}},\ \bibinfo {eid} {94} (\bibinfo {year}
  {2014})}\BibitemShut {NoStop}%
\bibitem [{\citenamefont {Yang}\ \emph {et~al.}(2007)\citenamefont {Yang} \emph
  {et~al.}}]{Yang:2007yr}%
  \BibitemOpen
  \bibfield  {author} {\bibinfo {author} {\bibfnamefont {X.}~\bibnamefont
  {Yang}} \emph {et~al.},\ }\href@noop {} {\bibfield  {journal} {\bibinfo
  {journal} {Astrophys. J.}\ }\textbf {\bibinfo {volume} {671}},\ \bibinfo
  {pages} {153} (\bibinfo {year} {2007})}\BibitemShut {NoStop}%
\bibitem [{\citenamefont {Lintott}\ \emph {et~al.}(2008)\citenamefont {Lintott}
  \emph {et~al.}}]{Lintott:2008ne}%
  \BibitemOpen
  \bibfield  {author} {\bibinfo {author} {\bibfnamefont {C.~J.}\ \bibnamefont
  {Lintott}} \emph {et~al.},\ }\href@noop {} {\bibfield  {journal} {\bibinfo
  {journal} {Mon. Not. R. Astron. Soc.}\ }\textbf {\bibinfo {volume} {389}},\
  \bibinfo {pages} {1179} (\bibinfo {year} {2008})}\BibitemShut {NoStop}%
\bibitem [{\citenamefont {{Hirata}}(2009)}]{2009MNRAS.399.1074H}%
  \BibitemOpen
  \bibfield  {author} {\bibinfo {author} {\bibfnamefont {C.~M.}\ \bibnamefont
  {{Hirata}}},\ }\href {\doibase 10.1111/j.1365-2966.2009.15353.x} {\bibfield
  {journal} {\bibinfo  {journal} {Mon. Not. Roy. Astron. Soc.}\ }\textbf
  {\bibinfo {volume} {399}},\ \bibinfo {pages} {1074} (\bibinfo {year}
  {2009})},\ \Eprint {http://arxiv.org/abs/0903.4929} {arXiv:0903.4929
  [astro-ph.CO]} \BibitemShut {NoStop}%
\bibitem [{\citenamefont {Singh}\ and\ \citenamefont
  {Mandelbaum}(2016)}]{Singh:2015sva}%
  \BibitemOpen
  \bibfield  {author} {\bibinfo {author} {\bibfnamefont {S.}~\bibnamefont
  {Singh}}\ and\ \bibinfo {author} {\bibfnamefont {R.}~\bibnamefont
  {Mandelbaum}},\ }\href {\doibase 10.1093/mnras/stw144} {\bibfield  {journal}
  {\bibinfo  {journal} {Mon. Not. Roy. Astron. Soc.}\ }\textbf {\bibinfo
  {volume} {457}},\ \bibinfo {pages} {2301} (\bibinfo {year} {2016})},\ \Eprint
  {http://arxiv.org/abs/1510.06752} {arXiv:1510.06752 [astro-ph.CO]}
  \BibitemShut {NoStop}%
\end{thebibliography}%

\end{document}